\documentclass[twocolumn, prx, aps, 10pt, superscriptaddress, longbibliography]{revtex4-2}      

\usepackage{silence}        
\usepackage{appendix}
\usepackage{graphicx,subfigure}                           
\usepackage[normalem]{ulem}                               
\usepackage[dvipsnames]{xcolor}                           
\usepackage{physics}                                      
\usepackage{dsfont}                                       
\usepackage{amsmath}									  
\definecolor{mycolor}{rgb}{0.61, 0.11, 0.19}
\definecolor{indigo}{rgb}{0.29, 0.0, 0.51}
\usepackage[
colorlinks=true, 
linkcolor=mycolor, 
citecolor=mycolor, 
urlcolor=mycolor]{hyperref}                                

\begin{document}
	
\title{Quantum Magic in Discrete-Time Quantum Walk}
\author{Vikash Mittal}
    \email{vikashmittal.iiser@gmail.com}
    \affiliation{Department of Physics, National Tsing Hua University, Hsinchu 300044, Taiwan}

\author{Yi-Ping Huang}
    \email{yphuang@phys.nthu.edu.tw} 
    \affiliation{Department of Physics, National Tsing Hua University, Hsinchu 300044, Taiwan}
    \affiliation{Physics Division, National Center for Theoretical Sciences, Taipei 10617, Taiwan}
    \affiliation{Institute of Physics, Academia Sinica, Taipei 115201, Taiwan}    

\begin{abstract}
Quantum magic, which accounts for the non-stabilizer content of a state, is essential for universal quantum computation beyond classically simulable resources. However, the way magic builds up during structured unitary dynamics remains largely open. Here, we investigate the generation and evolution of quantum magic in discrete-time quantum walks (DTQWs)--a simple, tunable unitary model realizable on a wide range of architectures. We use Stabilizer R\'enyi Entropy as a measure of quantum magic and investigate single- and two-walker quantum walks on a one-dimensional lattice, considering a wide range of initial coin states. Our results reveal that DTQWs can dynamically generate significant magic, with the amount and structure dependent on the initial state of the coin. In the case of a single walker, we found a nontrivial and complementary relationship between magic and entanglement at long times. In the two-walker setting, even states close to stabilizer form can evolve into highly magical states when subjected to quantum walk protocols. The generation of magic can occur independently of entanglement growth, emphasizing its distinct role as a quantum resource. We further show that the dynamical magic generation is robust under realistic noise—specifically, decoherence in the coin degree of freedom—demonstrating that this process persists across noisy settings. Finally, we discuss an experimentally feasible scheme to measure magic using current technology. Our findings position DTQWs as accessible and controllable platforms for producing quantum magic, offering a new perspective on their role in quantum information processing and reliable quantum computation.
\end{abstract}
    
\maketitle

\section{Introduction}
Quantum resources such as entanglement, coherence, and magic underpin the power of quantum technologies ~\cite{Horodecki2009, Brunner2014, Adesso2017, Gottesman1997, Gottesman1998}. Among them, quantum magic—the non-stabilizerness of a state—provides the computational leverage that lifts Clifford circuits beyond classical simulability ~\cite{Gottesman1997,Gottesman1998,Aaronson2004,Kitaev2005, Veitch2014, Kitaev2016, Gour2019, Winter2022}. Although static and circuit-based aspects of magic are well characterized, its dynamical emergence under coherent, experimentally controllable evolution remains largely unexplored~\cite{Veitch2014, Dai2022, Marcello2024}.

The question of how magic is generated, distributed, and controlled under structured, experimentally controllable unitary dynamics is still wide open. Previous studies, which are very few, have focused on magic state distillation~\cite{Kitaev2005}, random circuits or Clifford+T gate sets~\cite{Nicola2014}, which prioritize analytic tractability or complexity-theoretic classification, but often obscure the physical mechanisms that give rise to magic~\cite{Zhang2024, Piotr2025, Turkeshi2025, Rosario2024, Beri2024, Niroula2024, Turkeshi2024, Turkeshi2025}. In particular, the role of quantum interference, spatial structure, and coherent propagation in generating magic remains underexplored.

In this work, we bridge this gap by investigating the generation and the dynamics of quantum magic in discrete-time quantum walks~\cite{Aharonov1993, Kempe2003}—a minimal and widely studied model of unitary evolution with direct experimental relevance. DTQWs offer several key advantages for probing dynamical quantum resources. First, unlike Clifford circuits, which cannot generate magic from stabilizer states, generic DTQW dynamics are inherently non-Clifford, allowing magic to arise naturally even in single-particle regimes. Second, DTQWs preserve coherent interference and spatial correlations, providing a setting where the structure of magic generation can be directly linked to wave-like evolution. Third, they have been successfully implemented in a range of experimental systems, including photonic circuits~\cite{Regensburger2012, Schreiber2010, Schreiber2012}, trapped ions~\cite{Milburn2002, Schmitz2009, Karski2009}, and superconducting qubits~\cite{Pan2019, Pan2021, Yang2025} to name a few and offer continuous tunability through control parameters~\cite{Manouchehri2014}. While DTQWs are well-studied as generators of quantum resources like quantum entanglement~\cite{Vieira2013, Mittal2025}, quantum non-locality~\cite{Alberti2015, Angelo2019}, and quantum coherence~\cite{Kadian2021}, their capacity to generate quantum magic has remained unexplored. One of the very few studies along this direction is Ref.~\cite{QWMagic2025}, which investigates the generation of quantum magic in continuous-time multi-particle quantum walks.

Here we are interested in the following question: Can a very simple model of DTQW on a one-dimensional lattice generate a significant amount of quantum magic? We answer this question affirmatively by exploring both single- and two-particle DTQWs, initializing the coin state in various pure and mixed states, including stabilizer, non-stabilizer, and Werner-type states~\cite{Werner1989}. To quantify magic, we use the stabilizer R\'enyi entropy (SRE), a recently introduced and scalable measure of quantum magic suited for dynamical studies~\cite{Zhang2024, Piotr2025, Turkeshi2025}. In the single-walker case, we analyze how the time evolution of SRE depends on the initial coin state. We find that large initial SRE does not necessarily correlate with large asymptotic values, indicating that quantum walk dynamics can both enhance and suppress initial magic. Importantly, our analytical treatment of the long-time limit reveals a complementary relationship between entanglement and SRE in the system, supporting the view that magic constitutes a distinct resource. This suggests a potential trade-off, reminiscent of a monogamy-like relationship between the two resources. While a formal relation remains to be established, our findings highlight that entanglement and magic are complementary but fundamentally distinct aspects of quantum nonclassicality.

In the two-walker setting, we observe dynamical magic generation from a variety of initial states, including those with zero or high initial SRE. We highlight how interference-driven evolution transforms initially stabilizer states into complex, non-stabilizer configurations, and how the amount of magic generated can be quantitatively tuned by the initial configuration of the composite system. We also explore the dynamics of quantum magic when the two walkers are initialized in a Werner state, which interpolates between maximally entangled and separable configurations, providing a means to explore the full entanglement spectrum.

To assess practical relevance, we examine the robustness of magic generation against decoherence, incorporating a minimal coin dephasing model in the one-walker setting. We find that DTQWs maintain significant levels of magic over a wide range of noise strengths, demonstrating their viability as noise-resilient generators of quantum resources. Finally, we discuss an experimentally feasible protocol to measure magic using coin state tomography, followed by the measurements of the Pauli observables—well within reach of current technology available in DTQW platforms~\cite{Wang2020}.

Through these findings, we demonstrate that DTQWs serve as a robust, natural, and controllable generators of quantum magic. Their simplicity and accessibility for experiments make them ideal platforms for exploring the dynamics of magic. Our work contributes to the growing field of dynamical magic by providing new insights into how quantum advantages can arise and be manipulated through evolution, even within systems made up of elements that can otherwise be simulated classically.

The paper is organized as follows: In Section~\ref{sec:dtqw}, we introduce the DTQWs and set up the notations for the article along with the framework for quantifying quantum magic using Stabilizer R\'enyi Entropy. Sec.~\ref{sec:dtqw1} focuses on the dynamics of magic in single-walker systems, examining its dependence on initial coin states and behavior at very long times. In Sec.~\ref{sec:dtqw1_deco}, we introduce a very simple phase decoherence in the coin subspace to investigate the robustness of the quantum magic in noisy scenarios. In Sec.~\ref{sec:dtqw2}, we extend our results to two-walker systems, exploring how dynamics and initialization of the coin states influence the generation of magic. We also discuss a special case where the initial coin state is taken as a Werner state, revealing interesting dynamical features. We conclude in Sec.~\ref {sec:conclusion} with a summary of key results, their implications for quantum information processing, and the possible future research directions, along with some foundation aspects of our studies.

\section{Discrete-Time Quantum Walk}
\label{sec:dtqw}
Quantum walks, the quantum analog of classical random walks, have emerged as a versatile platform in quantum information science~\cite{Aharonov1993,Kempe2003,Nayak2000}. They exhibit rich interference patterns, allow for coherent control, and have been widely studied in the context of quantum algorithms~\cite{Ambainis2003,Childs2004,Shenvi2003}, quantum computation~\cite{Childs2009, Childs2013, Lovett2010}, and entanglement generation~\cite{Vieira2013, Mittal2025}. In both discrete-time and continuous-time frameworks, quantum walks have provided insights into the spread of quantum correlations and have been proposed for tasks ranging from search problems to quantum simulation~\cite{VenegasAndraca2012, Kadian2021}.

Here, we are interested in the discrete-time version of the quantum walk. A discrete-time quantum walk (DTQW) models the unitary evolution of a quantum particle, a walker, on a one-dimensional lattice, governed by the internal states of the walker, which are usually taken to be two-dimensional in the case of a one-dimensional system and referred to as a coin. The total Hilbert space is given by $\mathcal{H} = \mathcal{H}_C \otimes \mathcal{H}_P$, where $\mathcal{H}_C$ is the coin space spanned by $\{ \ket{\uparrow}, \ket{\downarrow} \}$, and $\mathcal{H}_P$ is the position space spanned by $\{ \ket{x} \mid x \in \mathds{Z} \}$. The evolution proceeds through repeated applications of a unitary step operator composed of a coin flip followed by a conditional shift. In this work, we consider a Hadamard coin
\begin{equation}
    H = \frac{1}{\sqrt{2}} 
\begin{pmatrix}
1 & 1 \\
1 & -1
\end{pmatrix},
\end{equation}
which acts on $\mathcal{H}_C$, creating an equal superposition between $\ket{\uparrow}$ and $\ket{\downarrow}$ states. The conditional shift operator $S$ moves the walker to the left or right depending on the state of the coin:
\begin{equation}
    S = \sum_{x \in \mathds{Z}} \dyad{\uparrow} \otimes \dyad{x+1}{x}  + \dyad{\downarrow} \otimes \dyad{x-1}{x},
\end{equation}
This operation entangles the coin and position degrees of freedom, introducing quantum interference into the walk dynamics. The unitary operator for one step of the walk is given by:
\begin{equation}
    U = S \cdot (H \otimes \mathds{1}_{P}),
\end{equation}
where $\mathds{1}_{P}$ is the identity operator on $\mathcal{H}_P$. For a given initial state of the composite system, $\ket{\Psi(0)}$, the state after $t$ time steps evolves as
\begin{equation}
    \ket{\Psi(t)} = U^t \ket{\Psi(0)} = \sum_{x \in \mathds{Z}} \sum_{\sigma = \uparrow, \downarrow} \Psi_x^{\sigma} \ket{\sigma} \otimes \ket{x}.
\end{equation}
where $\Psi_x^{\sigma}$ are the coefficients corresponding to the position $x$ and the coin state $\sigma = \{ \ket{\uparrow}, \ket{\downarrow} \}$. The total state at any time $t$ is a highly entangled state between position and coin degrees of freedom and cannot be factorized into the individual states of the subsystems. The coin operation under consideration is Hadamard, $H$, which belongs to the Clifford group. While the translation operator is non-local, we will see that this is the only part of the dynamics that is responsible for the nontrivial generation of several quantum resources in the system. 

While discrete-time quantum walks have been extensively studied for their entanglement properties, algorithmic applications, and transport behavior, their role in hosting or generating quantum magic has received \emph{no} attention. Please note here that when we say magic in quantum walk, it is the magic content in the state of the walker, which is a two-level system. Quantum magic refers to the non-stabilizer content of a quantum state—that is, the degree to which it departs from the set of stabilizer states. Since stabilizer states, along with Clifford operations, can be efficiently simulated on classical computers, the presence of magic is essential for achieving true quantum advantage in quantum computational processes. Quantifying magic is, therefore, central to understanding the usefulness of a quantum system for providing quantum advantage and enabling universal quantum computation.

\subsection{Stabilizer R\'enyi Entropy (SRE)}
The Stabilizer R\'enyi Entropy (SRE) is a quantifier of quantum magic, which measures the extent to which a quantum state deviates from the set of stabilizer states~\cite{Leone2022,Haug2023}. Stabilizer states form a subset of quantum states that can be described entirely by Clifford circuits and are efficiently simulable on classical computers~\cite{Gottesman1997,Gottesman1998}. Quantum states with non-zero magic, as captured by the SRE, represent a fundamental resource as they are essential for universal quantum computation and cannot be simulated on a classical machine.

Let $\rho$ be a quantum state describing $n$-qubits. The Stabilizer R\'enyi Entropy of order $\alpha$ is defined as~\cite{Leone2022}
\begin{equation} 
    \label{eq:SRE_alpha}
     \mathcal{M}_{\alpha}(\rho) = \dfrac{1}{1 -\alpha} \log \left( \dfrac{\sum_{P \in \mathcal{P}_n} \abs{\Tr(\rho P)}^{2 \alpha}}{\sum_{P \in \mathcal{P}_n} \abs{\Tr(\rho P)}^{2}} \right).
\end{equation}
where $\mathcal{P}_n$ are the $n$-qubit Pauli group string operators given by $\mathcal{P}_n = \{ \mathds{1}, \sigma_x, \sigma_y, \sigma_z\}^{\otimes n}$ with $\sigma_i$'s being the standard Pauli operators for a single qubit. The coefficients $c_n = \Tr(\rho P_n)$, along with an appropriate normalization, constitute a valid probability distribution. Then, the quantity defined above is the Rényi entropy associated with the probability of a quantum state when expanded in the basis of  Pauli strings.

For the current work, we use the Stabilizer R\'enyi entropy of order 2 for several reasons. First, it provides a computationally efficient measure with a closed analytical form for both pure and mixed states through Pauli expectation value summation. Recent theoretical work has rigorously established that SRE with $\alpha \ge 2$ are monotones for magic-state resource theory, making $\alpha=2$ the minimal order guaranteeing monotonicity~\cite{Leone2022, Leone2024}. Furthermore, SRE-2 is experimentally accessible via Pauli tomography~\cite{Leone2022, Oliviero2022, Kim2024}. Mathematically, it is expressed as
\begin{equation}
    \label{eq:renyientropy}
    \mathcal{M}(\rho) = -\log \left( \dfrac{\sum_{P \in \mathcal{P}_n} \abs{\Tr(\rho P)}^4}{\sum_{P \in \mathcal{P}_n} \abs{\Tr(\rho P)}^{2}} \right).
\end{equation}
and for a pure state $\ket{\Psi}$, it reduces to 
\begin{equation}
    \mathcal{M}(\rho) = -\dfrac{1}{2^n}\log \left(  \sum_{P \in \mathcal{P}_n} \abs{\expval{P}{\Psi}}^4  \right).
\end{equation}
The SRE has several desirable properties: it is invariant under Clifford operations, additive under tensor products, and non-increasing under stabilizer-preserving operations~\cite{Leone2022}. These characteristics make it particularly well-suited for analyzing magic generation in dynamical systems such as quantum circuits~\cite{Piotr2025, Turkeshi2025}.

In Ref.~\cite{Wang2023,Fuchs2024,Zhewei2025}, the upper-bound of SRE of order $\alpha$, for a $d$-dimensional Hilbert space, was provided which reads
\begin{equation}
    \label{eq:upperbound}
    \mathcal{M}_{\alpha}(\ket{\psi}) \le \dfrac{1}{1 - \alpha} \log \dfrac{1 + (d-1)(d+1)^{1 - \alpha}}{d}.
\end{equation}
So, for $\alpha = 2$ and for single- and two-qubit systems, this bound results in $\mathcal{M}_2(\ket{\psi}) \le \log(3/2) \approx 0.4055$ and $\mathcal{M}_2(\ket{\psi}) \le \log(5/2) \approx 0.9163$ respectively. We should note here that these bounds can be achieved (if possible) for pure quantum states in $d$-dimensional Hilbert space.

\section{DTQW with one walker}
\label{sec:dtqw1}
\begin{figure}
    \centering
    \subfigure{\includegraphics[width=0.47\textwidth]{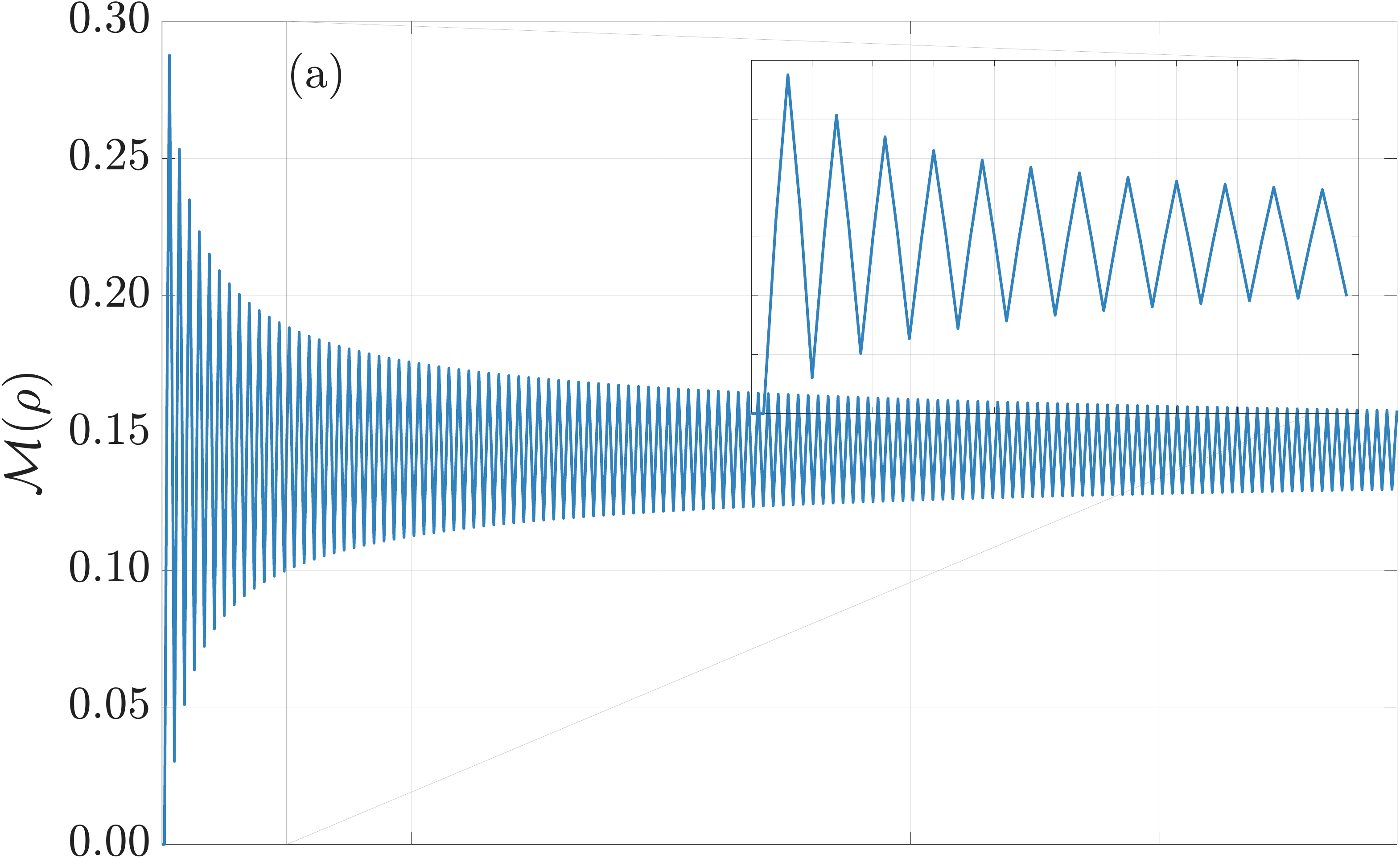}}
    \subfigure{\includegraphics[width=0.47\textwidth]{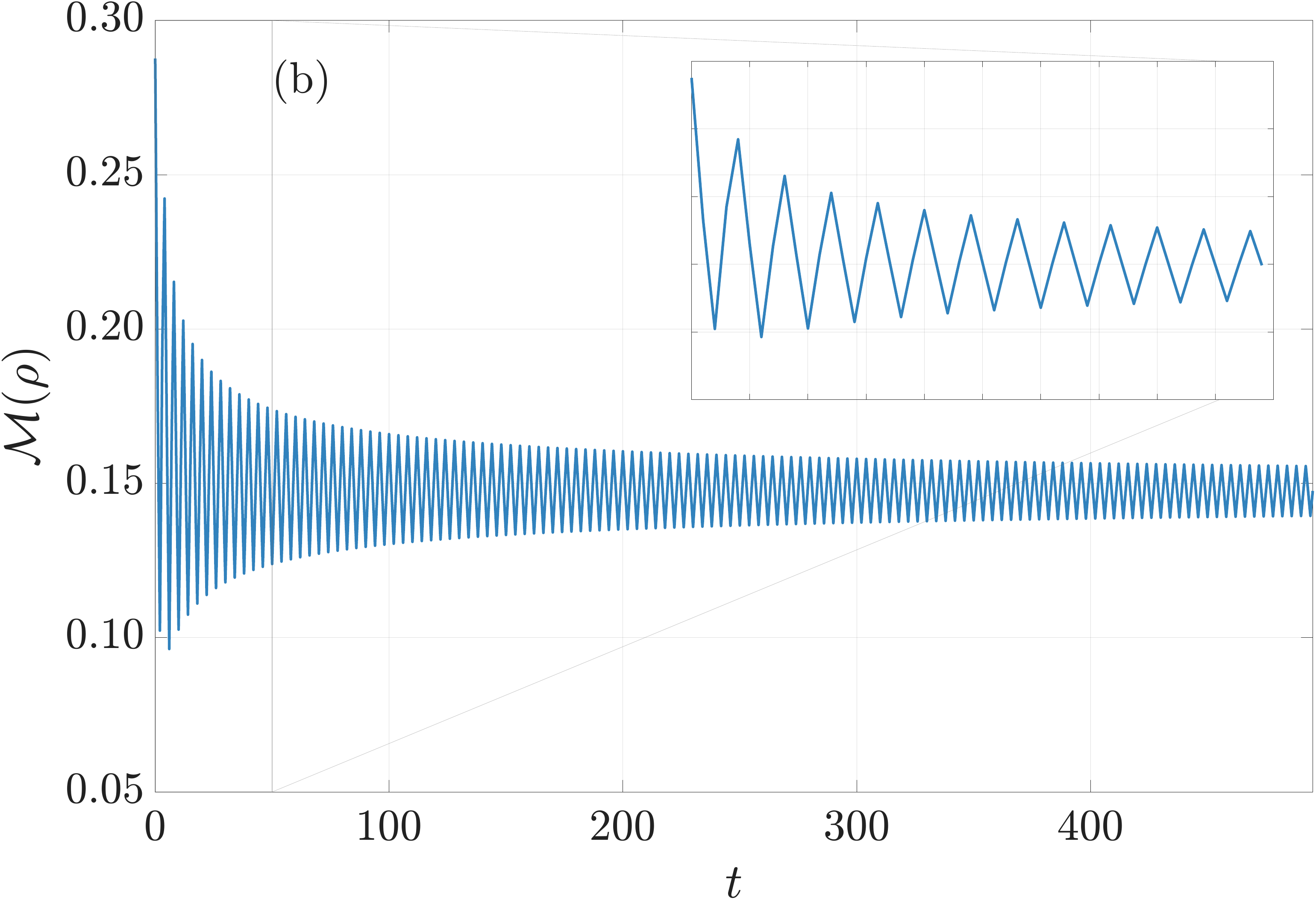}}
    \caption{The SRE as a function of time for the initial state of coin to be (a) $\ket{\Psi(0)}_C = \ket{\uparrow}$, (b) $\ket{\Psi(0)}_C = (\ket{\uparrow} + e^{i \pi/4} \ket{\downarrow})/\sqrt{2}$ while lattice part is localized in both the cases at $\ket{x = 0}$. The plots in the inset show the growth of non-stabilizerness at initial time steps. The system size is taken to be $1001$ for both plots.}
    \label{fig:fixedtheta_t}
\end{figure}

\begin{figure*}
    \centering
    \subfigure[]{\includegraphics[height=0.3\textwidth]{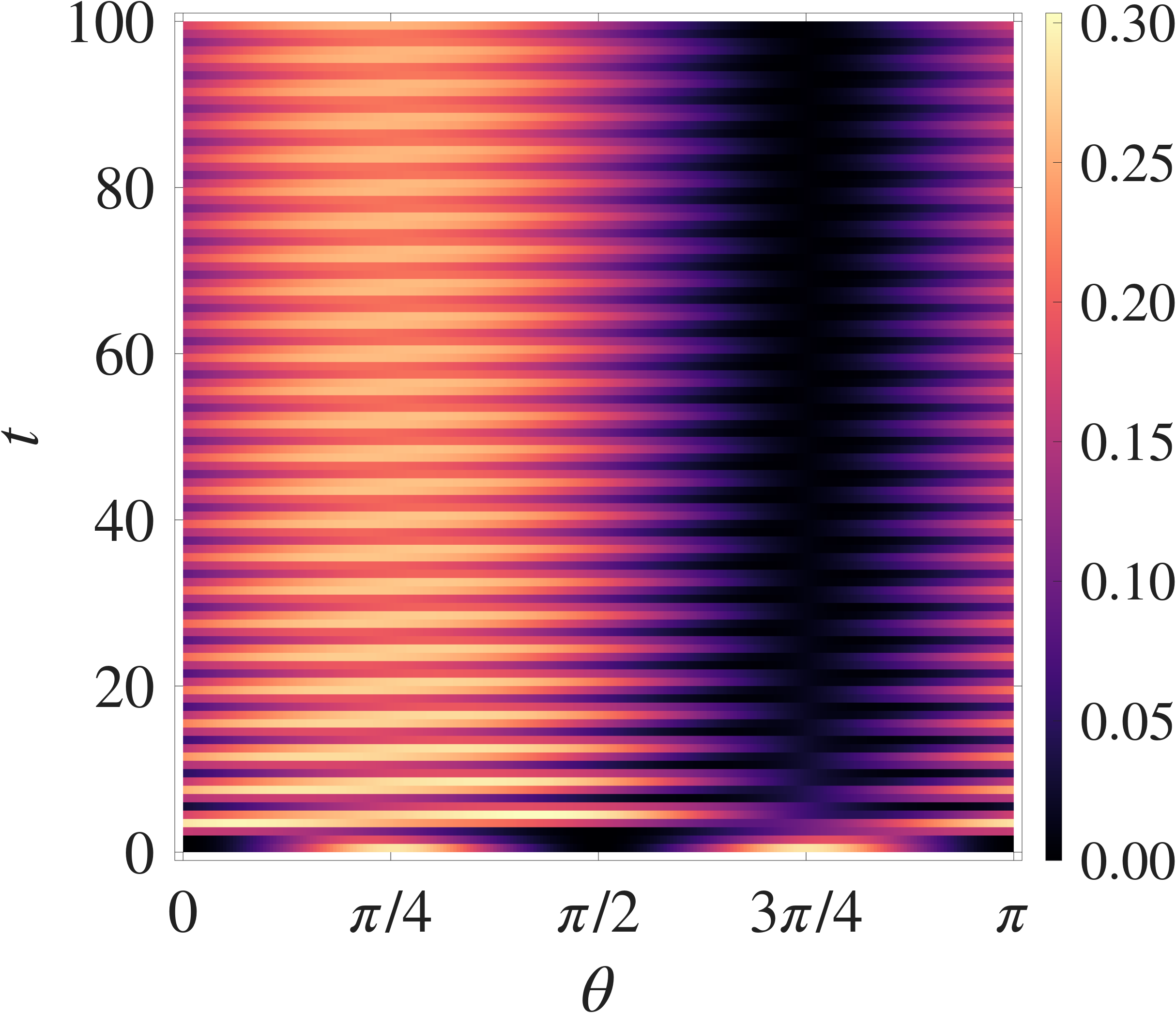}}
    \subfigure[]{\includegraphics[height=0.3\textwidth]{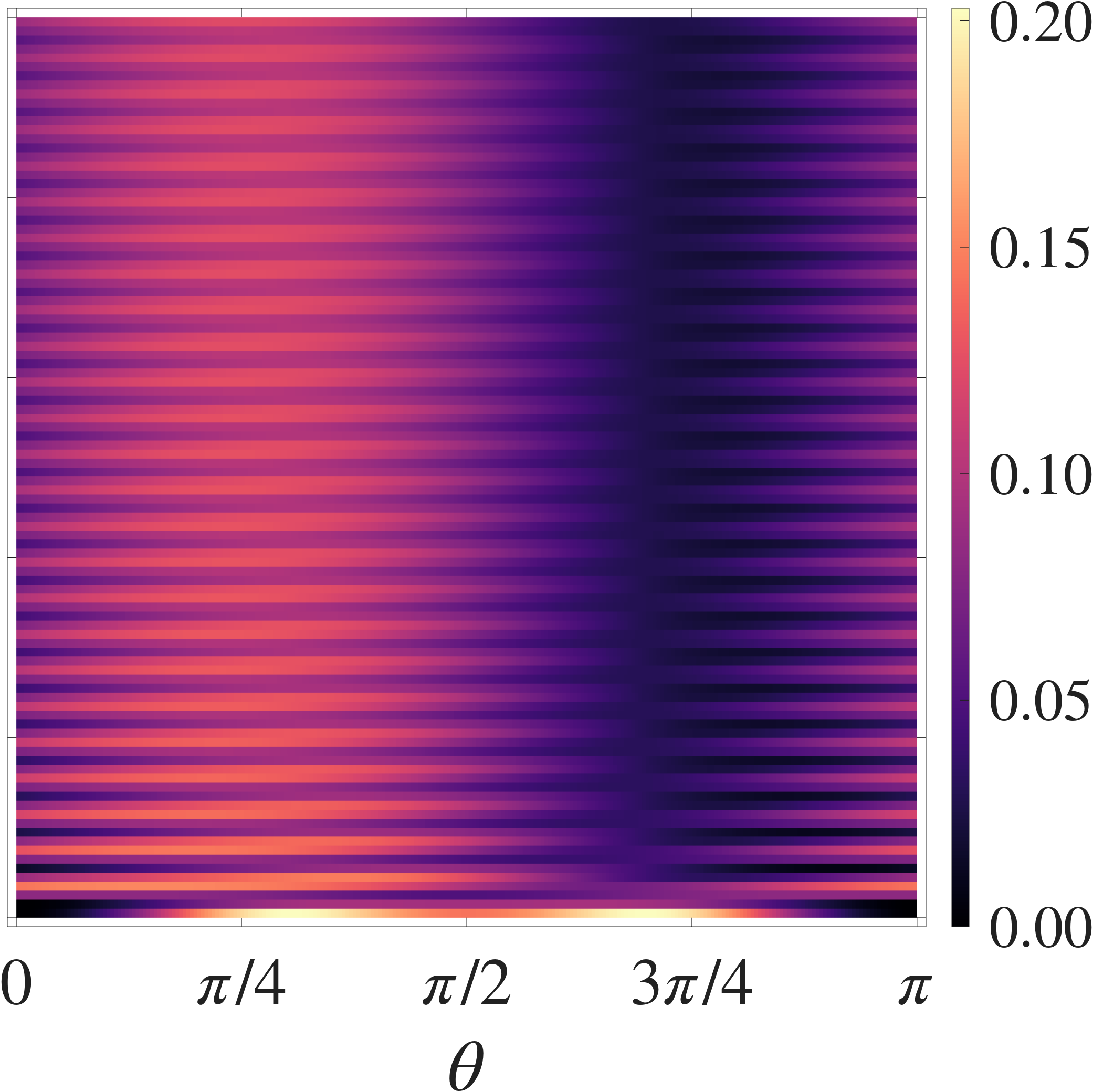}}
    \subfigure[]{\includegraphics[height=0.3\textwidth]{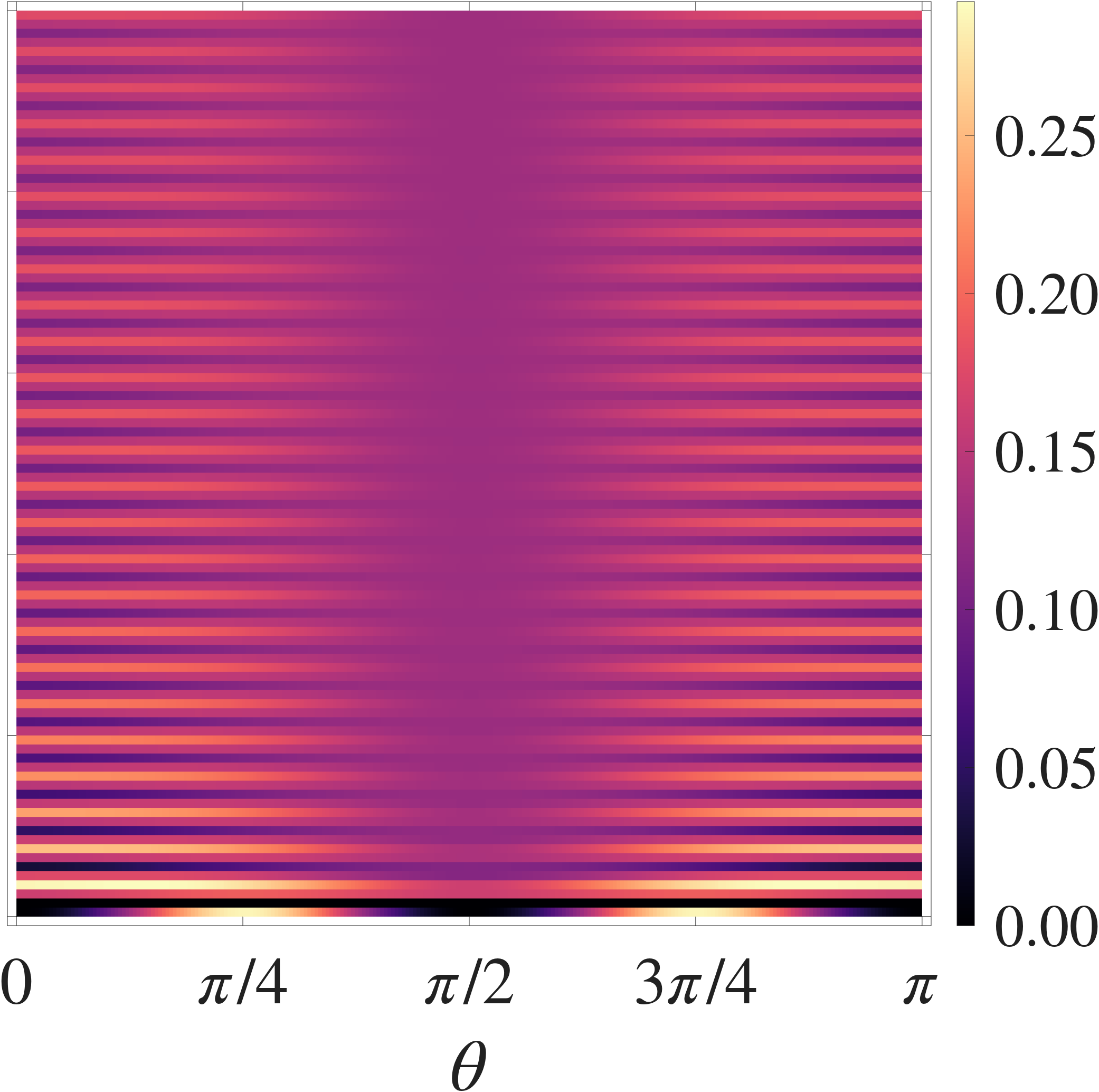}}
    \caption{The SRE as a function of time and the parameter $\theta$ in the initial state of the coin for (a) $\phi = 0$, (b) $\phi = \pi/4$, (c) $\phi = \pi/2$. We can see the oscillating patterns in the density plots for almost all the values of the parameters $\theta$ and $\phi$; however, they have different amplitudes and oscillation rates. The system size is taken to be $1001$ for all the plots.}
    \label{fig:theta_t}
\end{figure*}
\begin{figure*}
    \centering
    \includegraphics[width=0.95\textwidth]{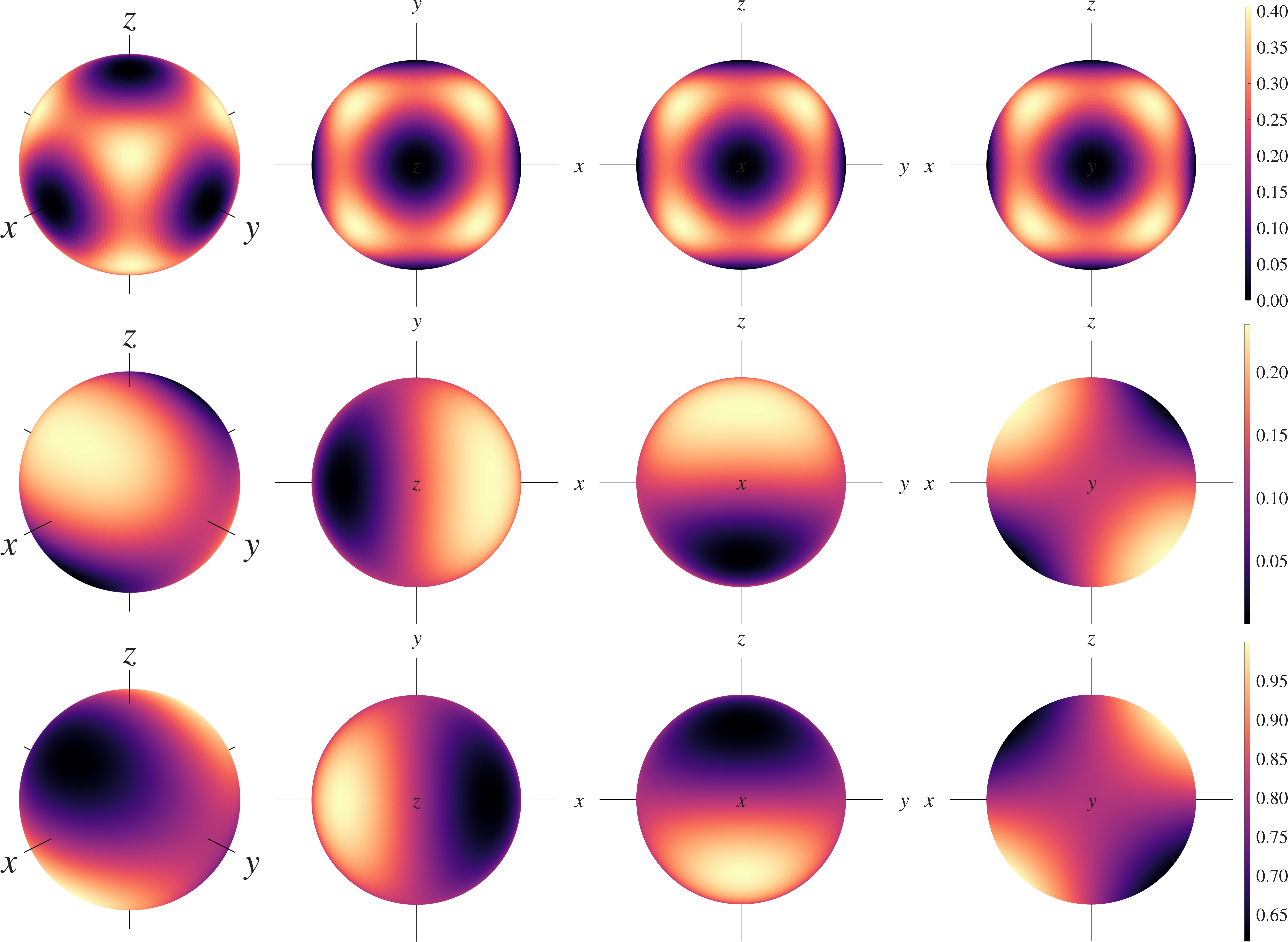}
    \caption{Top Row: The SRE for different initial coin states, parameterized using the Bloch sphere representation, at $t = 0$, and in Middle $\&$ Bottom row, the SRE and von-Neumann entanglement entropy in the asymptotic limit as a function of the $\theta$ and $\phi$ from Eq.~\eqref{eq:initialstate}, respectively. We also plot the projection of the Bloch sphere in $x-y$, $y-z$, and $z-x$ planes for clarity. We observe a complementary behavior of SRE and the von-Neumann entropy in the long-time limit.}
    \label{fig:large_t}
\end{figure*}
Building upon the framework of discrete-time quantum walks and the utilization of Stabilizer R\'enyi Entropy as a quantifier of quantum magic, we are now ready to focus on the dynamics of magic generation within a single-walker context. The aim is to analyze how a quantum walker, initialized in a general coin superposition state and when subjected to unitary evolution on a one-dimensional lattice, generates and accumulates magic over time.

We assume that the walker is initially localized at the origin of the lattice, with the coin represented by an arbitrary pure state, parametrized using the Bloch sphere. The initial state of the total system thus reads  
\begin{equation}
    \ket{\Psi(0)} = \left( \cos(\theta/2) \ket{\uparrow} + e^{i \phi} \sin(\theta/2) \ket{\downarrow} \right) \otimes \ket{x = 0}
    \label{eq:initialstate}
\end{equation}
where $\theta \in [0, \pi]$ and $\phi \in [0, 2\pi]$. Such systematic consideration of all arbitrary initial states for the coin in our analysis effectively captures all possible single-qubit coin operations, as any initial state can be reached by a suitable unitary~\cite{Renato2013}.

We evolve the quantum walk with the time evolution operator $U$, and the total density matrix at time $t$ is written as
\begin{equation}
    \rho(t) = U^t \rho(0) (U^{\dagger})^t.
\end{equation}
At each time step, we analyze the reduced coin state $(\rho_C(t))$ obtained by tracing out the position degree of freedom, and compute the SRE given in Eq.~\eqref{eq:renyientropy} to quantify the amount of magic generated in the coin subsystem. Note that the SRE formulation naturally extends to mixed states using an appropriate normalization factor as given in Eq.~\eqref{eq:renyientropy}, maintaining essential properties including invariance under Clifford operations and monotonicity under stabilizer-preserving maps~\cite{Leone2022}.

In Fig.~\ref{fig:fixedtheta_t}, we plot the time evolution of Stabilizer R\'enyi Entropy (SRE) for the walker localized at $\ket{x = 0}$ and two distinct initial coin states: (a) a pure state $\ket{\uparrow}$ and (b) a superposed state $ \ket{\Psi(0)}_C = (\ket{\uparrow} + e^{i \pi/4} \ket{\downarrow})/\sqrt{2}$. The difference between the two initial states is in the initial amount of quantum magic. The state $\ket{\uparrow}$ initially has zero magic, while the other choice of initial state has finite magic initially. In Fig.~\ref{fig:fixedtheta_t}(a), the SRE remains zero after one step and shows a rapid increase after that, which is followed by a saturation, suggesting that quantum magic accumulates quickly and stabilizes as the walk proceeds. The zero magic after the first step can be ascribed to the fact that the state after the first step of the quantum walk is a Bell-like state and is maximally entangled. In the other case shown in Fig.~\ref{fig:fixedtheta_t}(b), the amount of magic is suppressed by the dynamics, and we again observe a saturation as time increases. In both cases, we see rapid oscillations in the dynamics, with the amplitude of the oscillations also sensitive to the initial state. The probability of finding the walker at odd (even) lattice sites vanishes for even (odd) time steps, and this could be a possible reason behind the oscillatory behavior of the magic with time. We observe similar oscillations in the dynamics of entanglement in a discrete-time quantum walk~\cite{Carneiro2005}. Here, we should clarify that the oscillations shown in Fig.~\ref{fig:fixedtheta_t} are those of the SRE, which quantifies the non-stabilizer (or the magic) content of the coin state, not its von Neumann entropy. While von Neumann entropy tracks information or entanglement between coin and position—oscillating persistently around a constant average—SRE measures non-Clifford resource generation and depletion. These two quantities are fundamentally different. The higher final SRE in the superposition case highlights the crucial role of initial coin coherence in facilitating magic generation. This indicates that superposed coin states are more effective at populating non-stabilizer subspaces over time.

Further, in Fig.~\ref{fig:theta_t}, we investigate the SRE as a function of both time and the Bloch sphere angle $\theta$ for fixed values of $\phi$ as: (a) $\phi = 0$, (b) $\phi = \pi/4$, and (c) $\phi = \pi/2$. The time dynamics of the quantum walks indicate that the generation of magic depends in a nontrivial manner on the choice of the initial coin state. For $\phi=0$, the SRE shows a strong dependence on $\theta$, peaking at intermediate values. If we look at the right side of Fig.~\ref{fig:theta_t}(a), we see a black region which corresponds to a lack of magic in the system and indicates that for some values of $\theta$, the system has a considerable amount of magic initially, but it decays very rapidly. As the $\phi$ increases, the regions of high magic shift and broaden, indicating increased sensitivity of magic generation to initial phase coherence. For $\phi = \pi/2$, the density plot suggests periodic dependencies on $\theta$ and underlines the constructive role of coin state superposition. Another non-trivial feature to note in the case of $\phi = \pi/2$ is that the magic after the first step is zero and peaks at the second time-step irrespective of the $\theta$ value. 

\subsection{SRE at asymptotic time}
Even for our simple setting of a single walker with a fixed coin operator, closed-form analytical expressions are not available except for the long-time limit, due to the fundamental complexity of quantum interference across exponentially many paths~\cite{Nayak2000}.

So, we now investigate the saturation levels of SRE at the asymptotic time limit for a fixed localized position of the walker and different initializations of the coin state.   By following the calculations of Ref.~\cite{Nayak2000,Abal2006,Abal2006a}, we can write the reduced density matrix of the coin at large time as
\begin{equation}
    \lim_{t \rightarrow \infty} \rho_C = \begin{pmatrix}
        A & B \\
        B^* & C
    \end{pmatrix}
    \label{eq:reduced_coin}
\end{equation}
where 
\begin{equation}
  \begin{gathered}
      C = b_1 \left(3 + \sqrt{2} - 2\sin^2\theta + \sin(2\theta)\cos\phi \right) \\
    B = b_1 \left(1 - 2\sin^2\theta + \sin(2\theta)(-\cos\phi + i\sqrt{2}\sin\phi) \right)
  \end{gathered}
\end{equation}
with 
\begin{equation}
    b_1 = \frac{\sqrt{2} - 1}{2\sqrt{2}}.
\end{equation}
The trace constraint leads to $A + C = 1$. Now, using the reduced state of the coin given in Eq.~\eqref{eq:reduced_coin}, we can estimate the SRE at asymptotic time, and the results are plotted in Fig.~\ref{fig:large_t}. In the top row of Fig.~\ref{fig:large_t}, we plot SRE on the Bloch sphere~\cite{Nielsen_Chuang_2010} for all the possible values of $\theta$ and $\phi$ in the allowed range at $t = 0$ and we can identify \emph{exactly} eight states which saturate the bound given by Eq.~\eqref{eq:upperbound}. In the Middle and the Bottom row of Fig.~\ref{fig:large_t}, we compare SRE and von Neumann entanglement entropy (for the reduced density matrix of the coin), respectively, at asymptotic time across a range of $\theta$ and $\phi$. In the subsequent columns, we also plot the projection of the Bloch sphere in $x-y$, $y-z$, and $z-x$ planes that provides a better representation of the behavior of the two quantities for different parameters. The results show that high initial SRE does not always lead to high asymptotic values, implying that the dynamics of quantum walk can both enhance and suppress initial magic. 

The oscillations reported here are in the stabilizer R\'enyi entropy (SRE), not the von Neumann coin entropy classically reported for DTQWs. While entropy quantifies information content, SRE quantifies computational non-stabilizer resources. We observe regimes where SRE and entropy oscillate out of phase, underscoring different physical mechanisms.

Our results reveal an interesting and nontrivial trade-off between quantum magic and coin-position entanglement in discrete-time quantum walks (DTQW) involving a single walker. When entanglement entropy reaches its maximum, the reduced coin state tends to be closer to the stabilizer polytope, resulting in minimal magic, and vice versa. Strong entanglement operationally "delocalizes" non-Clifford features into nonlocal correlations, while high local non-stabilizerness suppresses the ability to establish maximal bipartite entanglement. Although this phenomenon resembles a monogamy relation, it is more accurately characterized as a complementarity or a “monogamy-style trade-off” between two distinct resources. 

Formalizing this behavior within a unified resource-theoretic framework—potentially through an inequality that constrains the sum of entanglement and magic—remains an intriguing open problem. Addressing this could yield deeper insights into the interplay of quantum resources under unitary dynamics.

This complementary pattern between the two resources suggests that while both are forms of non-classical correlations, they may not be simultaneously maximized. This supports the idea that magic captures a distinct type of resource compared to entanglement.

\section{Decoherence in DTQW} 
\label{sec:dtqw1_deco}
\begin{figure}
    \centering
    \includegraphics[width=0.45\textwidth]{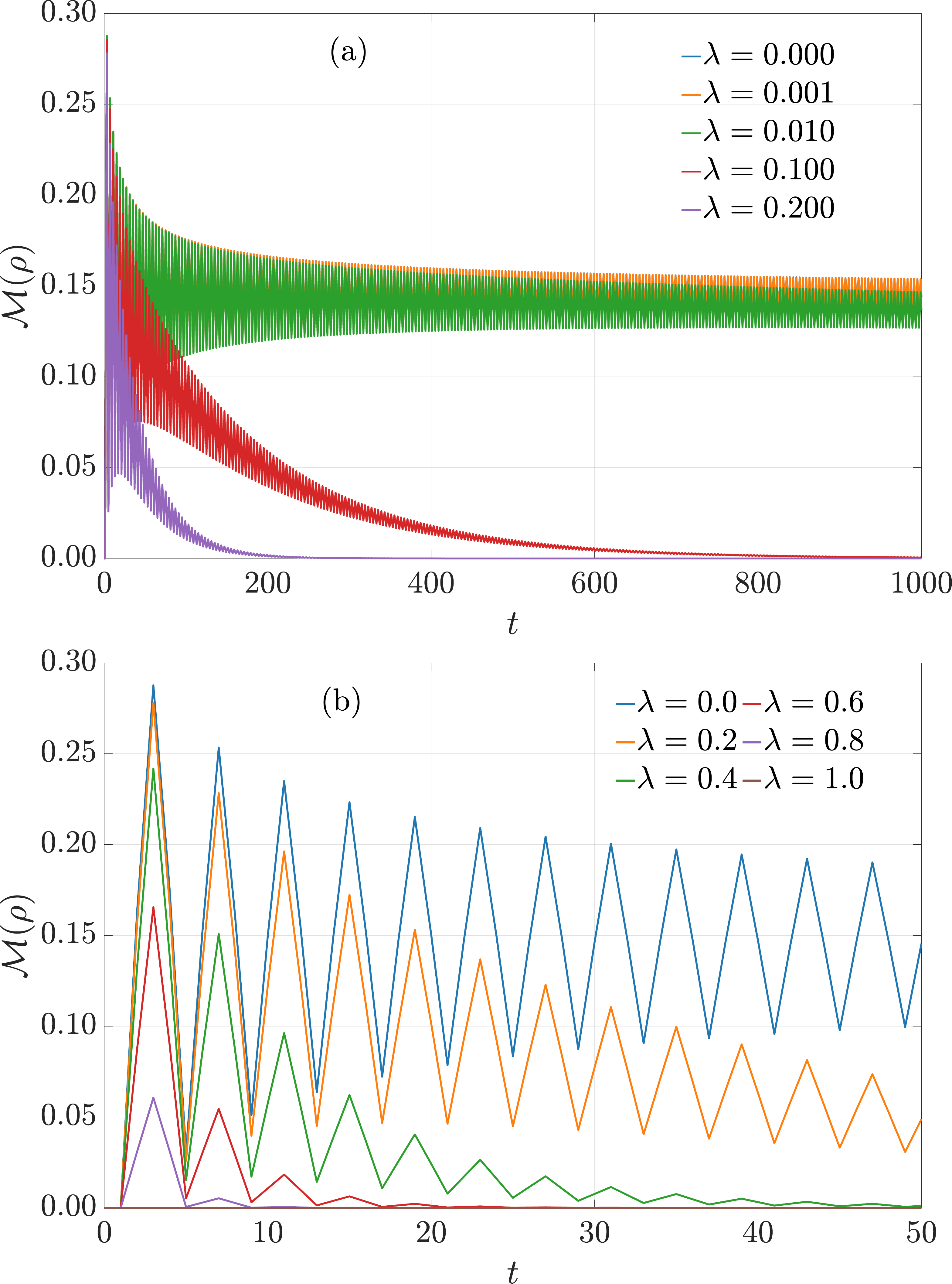}
    \caption{The time evolution of the SRE for a fixed initial coin state, $\ket{\Psi(0)}_C = \ket{\uparrow}$, under varying decoherence strengths $\lambda$. (a) Long-time behavior of $\mathcal{M}(\rho)$ up to $t = 1000$ for weak decoherence values. A small but nonzero values of $\lambda$ lead to gradual damping of oscillations, while strong decoherence suppresses magic more rapidly (b) Short-time dynamics of $\mathcal{M}(\rho)$ for a broader range of decoherence strengths. While strong decoherence quickly kills magic, weak and moderate decoherence still allow nontrivial oscillations and a delayed decay of magic, indicating robustness at early times. The lattice part is localized at $\ket{x = 0}$.}
    \label{fig:1D_Deco}
\end{figure}
\begin{figure*}
    \centering
    \includegraphics[width=\textwidth]{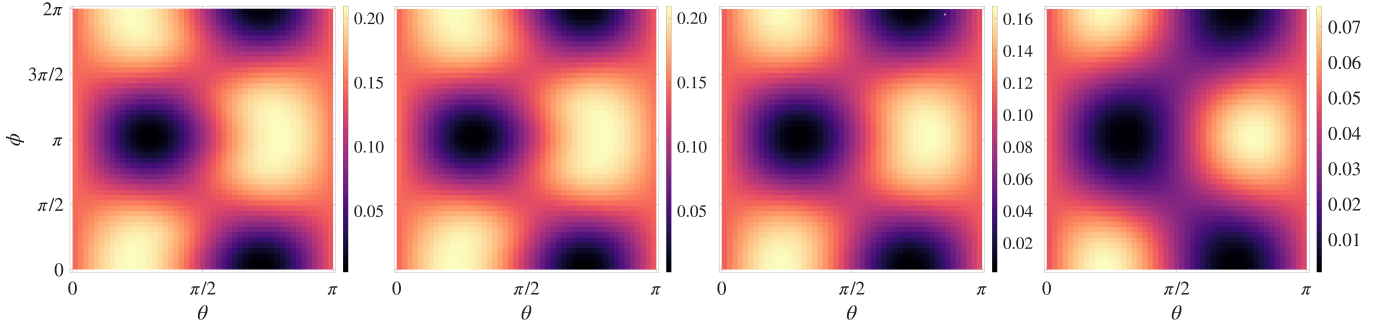}
    \caption{Stabilizer R\'enyi Entropy $\mathcal{M}(\rho)$ at time $t = 50$ as a function of the initial coin state parameterized by $\theta \in [0, \pi]$ and $\phi \in [0, 2\pi)$ for decoherence strengths $\lambda = 0$, $0.01$, $0.10$, and $0.20$. While the maximum SRE decreases with increasing $\lambda$, the overall landscape pattern remains qualitatively similar across these cases, indicating robustness in the structure of magic generation. For example, for $\lambda = 0.1$ we see a reduction of only $\approx 20 \%$ in the peak value of the SRE.}
    \label{fig:3D_Deco}
\end{figure*}
\begin{figure}
    \centering
    \includegraphics[width=0.45\textwidth]{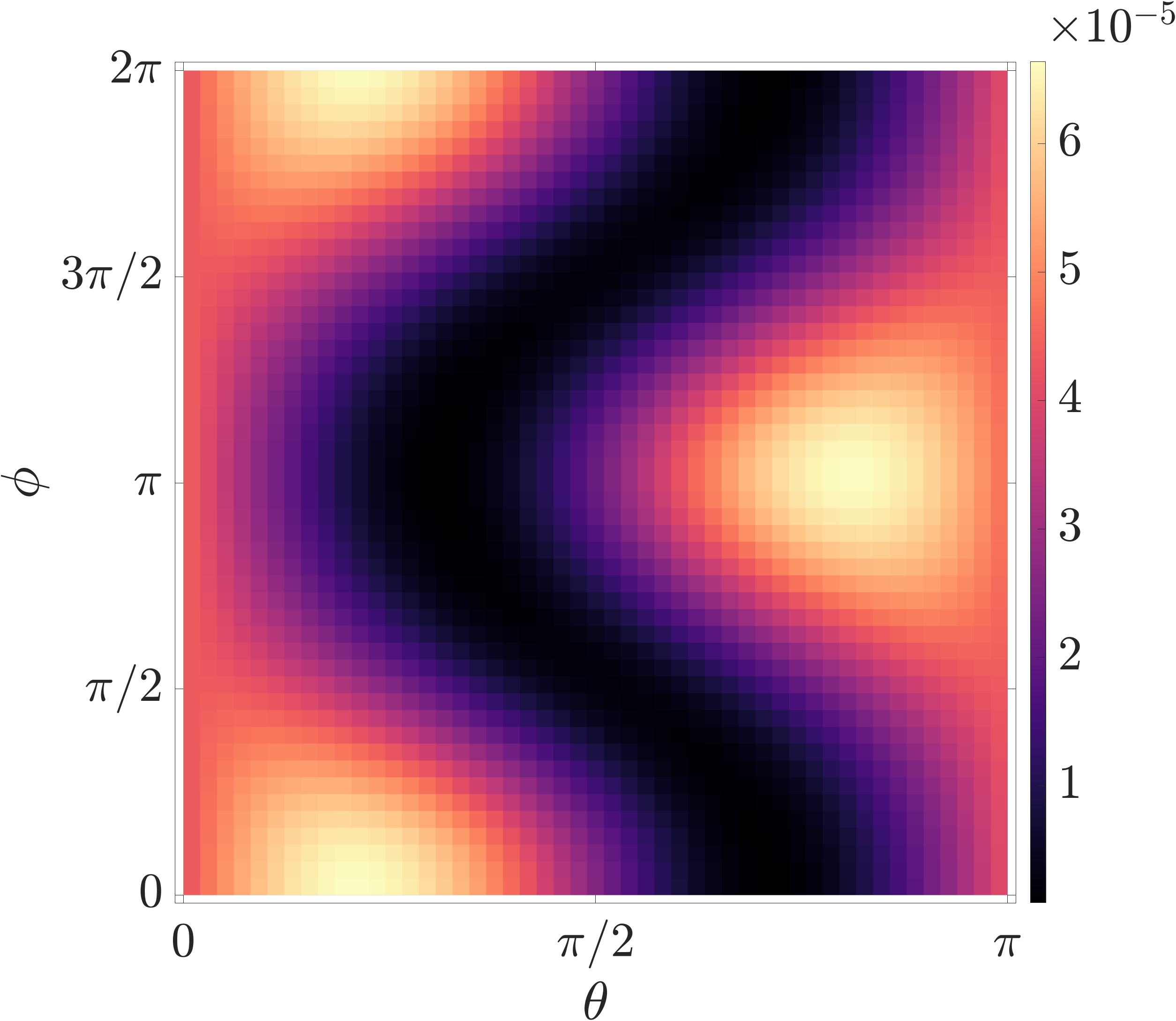}
    \caption{Stabilizer R\'enyi Entropy $\mathcal{M}(\rho)$ at time $t = 50$ as a function of the initial coin state parameterized by $\theta \in [0, \pi]$ and $\phi \in [0, 2\pi)$ for $\lambda = 0.50$, where the magic is suppressed almost across the various initial configurations. The maximum SRE in this case is $\sim 6.5 \times 10^{-5}$, indicating the near-total fragility of magic under strong decoherence.}
    \label{fig:3D_Deco2}
\end{figure}
In the previous section, we have seen that the unitary evolution of discrete-time quantum walks can generate a significant amount of quantum magic in the coin subspace. However, its practical utility depends critically on its performance and fidelity under noisy settings. In realistic quantum systems, interactions with the environment are unavoidable, which are unwarranted for quantum resources, including quantum magic. Therefore, to evaluate the resilience of magic in our setup, we introduce a minimal decoherence model and study its effect on the generation and persistence of quantum magic in the single-walker quantum walk.

Since our primary aim is to investigate the magic content specifically in the coin subspace, it is natural to examine how decoherence affects this same subspace. Thus, we implement a simple yet insightful model of decoherence in the coin space via dephasing-type measurements. At each time step, the coin undergoes a partial projective measurement in the $\sigma_z$ basis, characterized by the two measurement operators~\cite{Nielsen_Chuang_2010}
\begin{equation}
M_{\pm} = \dfrac{1}{2} (\mathds{1} \pm \lambda \sigma_z ),
\end{equation}
where $\lambda \in [0, 1]$ quantifies the strength of the decoherence. When $\lambda = 0$, the operators reduce to $M_{\pm} = \frac{1}{2} \mathds{1}$, corresponding to no decoherence, while $\lambda = 1$ corresponds to full projective measurement in the computational basis and washes away all the quantum coherence in the system.

The associated Kraus operators that govern the quantum channel are taken as the positive square roots of the measurement operators:
\begin{equation}
E_{\pm} = \sqrt{M_{\pm}}.
\end{equation}
These operators act on the coin degree of freedom and model weak measurement or dephasing, depending on the strength $\lambda$.

The effect of this channel is incorporated in the dynamics of the DTQW by utilizing the full density matrix of the composite system. The full density matrix of the system $\rho(t)$ evolves according to~\cite{Nielsen_Chuang_2010}
\begin{equation}
    \rho(t) = \sum_{i = \pm} (E_i \otimes \mathds{1}_P) U \rho(0) U^{\dagger} (E_i^{\dagger} \otimes \mathds{1}_P)
\end{equation}
where $E_{\pm}$ are the Kraus operators. This evolution is iterated over multiple steps, and the reduced coin state is obtained at each time by tracing over the position degrees of freedom. We then compute the SRE of the coin state to assess how magic is affected by the presence of decoherence.

To investigate the dynamical stability of quantum magic under the decoherence model explained above, we compute the time evolution of the $\mathcal{M}(\rho)$ of the reduced coin state for a fixed initial condition and various values of the decoherence strength $\lambda$. The results are shown in Fig.~\ref{fig:1D_Deco}, with Fig.~\ref{fig:1D_Deco}(a) covering a long time scale ($t \leq 1000$) and Fig.~\ref{fig:1D_Deco}(b) zooming into the short-time dynamics ($t \leq 50$) for more detailed investigation.

In Fig.~\ref{fig:1D_Deco}(a), we observe that in the absence of decoherence ($\lambda = 0$), the magic exhibits persistent oscillations around a stable average value, indicating sustained non-stabilizer content in the system. As decoherence is introduced, even at extremely small values ($\lambda = 0.001$ and $0.010$), the oscillations become damped over time, but the system still retains a significantly large amount of steady-state magic. As we further increase the decoherence strength ($\lambda = 0.100$), the damping becomes more significant, and the magic decays gradually towards a lower asymptotic value. At stronger decoherence ($\lambda = 0.200$), we observe a rapid suppression of magic, with the SRE decaying to near zero over a few hundred steps. However, we observe a presence of a significant amount of SRE for times $t \le 50$.

So, we focus on the short-time scale for a more detailed investigation of the behavior of the quantum magic at early times, which is particularly important for understanding the robustness of magic over experimentally relevant timescales. For weak to moderate decoherence strengths ($\lambda = 0.2$, $0.4$), the magic initially grows and exhibits oscillatory behavior similar to the unitary case ($\lambda = 0$), although with visibly reduced amplitude. These oscillations persist over several tens of time steps, suggesting that magic generation is relatively robust against moderate decoherence in the short term. For larger values of $\lambda$ ($0.6$, $0.8$, and $1.0$), the magic quickly decays to zero after just a few steps, indicating that strong decoherence effectively suppresses non-stabilizer features almost immediately.

Together, these observations indicate that quantum magic generated via DTQWs is moderately robust to weak and even intermediate levels of dephasing-like decoherence, especially over short to intermediate timescales. This is of particular importance because it makes quantum magic generated in the DTQW setting a usable resource in near-term quantum devices operating under realistic noise levels, provided that decoherence can be sufficiently controlled.

To further investigate the sensitivity of magic generation to the initial coin state under decoherence, we compute SRE $\mathcal{M}(\rho)$ at a fixed time $t = 50$ for a full range of pure coin states parametrized on the Bloch sphere and is given in Eq.~\eqref{eq:initialstate}.

We present results in Fig.~\ref{fig:3D_Deco} and Fig.~\ref{fig:3D_Deco2} for decoherence strengths $\lambda = 0$, $0.01$, $0.10$, $0.20$, and $0.50$ respectively. As expected, the introduction of decoherence leads to a progressive suppression of the generated magic. However, a notable observation is that the overall pattern or structure of the magic landscape remains qualitatively intact for $\lambda \leq 0.20$. That is, the regions of the Bloch sphere that yield relatively higher or lower magic in the unitary case continue to exhibit the same relative behavior under weak and moderate decoherence.

The maximum values of $\mathcal{M}(\rho)$ across the Bloch sphere drop significantly with increasing $\lambda$. Specifically, the peak SRE values for $\lambda = 0$, $0.01$, $0.10$, and $0.20$ are approximately: $ \mathcal{M}_{\text{max}} = 0.20,\ 0.20,\ 0.16,\ \text{and } 0.07,$ respectively. These values indicate that while very weak decoherence ($\lambda = 0.01$) has little effect on the maximum achievable magic, stronger noise rapidly diminishes the capacity of the system to generate high-magic states.

To further illustrate the fragility of magic under strong decoherence, we plot SRE for $\lambda = 0.50$ in Fig.~\ref{fig:3D_Deco2}, where the observed maximum value of $\mathcal{M}(\rho)$ drops to just $6.5 \times 10^{-5}$, effectively vanishing for all practical purposes. At this strength, the quantum walk is no longer capable of maintaining any significant non-stabilizer character, regardless of the initial coin state.

These observations confirm that while quantum magic in DTQWs is initially robust across a broad class of initial conditions, it becomes increasingly fragile under stronger decoherence. Still, the structural persistence of the landscape under moderate noise suggests that controlled experimental environments may preserve useful levels of magic for a meaningful duration of the dynamics.

\section{DTQW with two walkers}
\label{sec:dtqw2}
\begin{figure}
    \centering
    \includegraphics[width=0.45\textwidth]{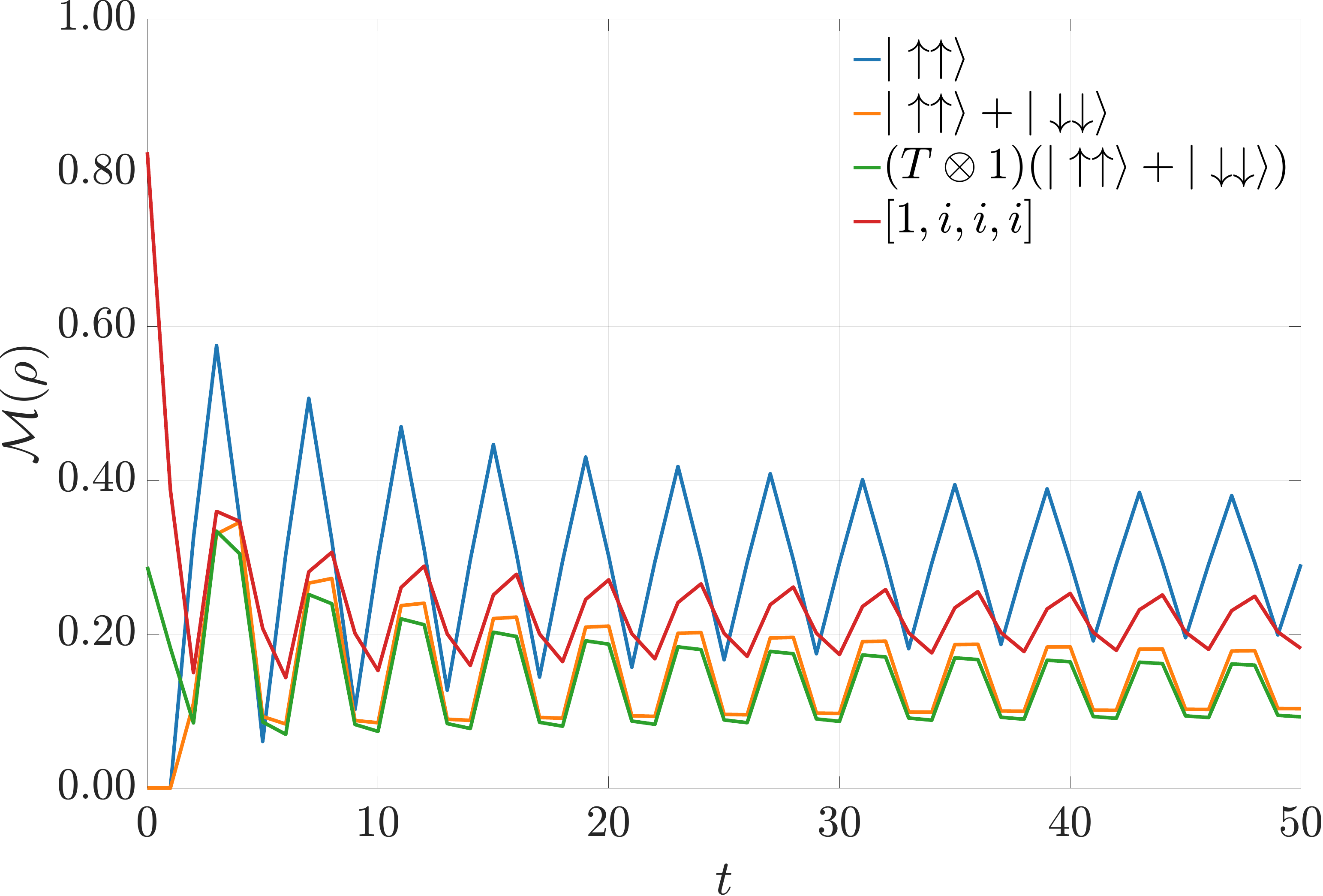}
    \caption{The SRE as a function of time for the different initialization of the coins at $t = 0$. Here, $T$ is the single-qubit magic gate and $[1, i, i, i]$ represents  $\ket{\chi} = \left( \ket{\uparrow \uparrow} + i \ket{\uparrow \downarrow} + i\ket{\downarrow \uparrow} + i\ket{\downarrow \downarrow} \right)/2$, a two-qubit state with maximal amount of magic~\cite{Zhewei2025}. Surprisingly, we see that an entangled state, a magic-enhanced state, and a state with initially maximal magic show limited growth in magic over time, while a simple product state results in significantly higher SRE. The system size is taken to be $101 \times 101$.}
    \label{fig:fixedtheta_t_twowalkers}
\end{figure}
\begin{figure*}
    \centering
    \subfigure[]{\includegraphics[width=0.3\textwidth]{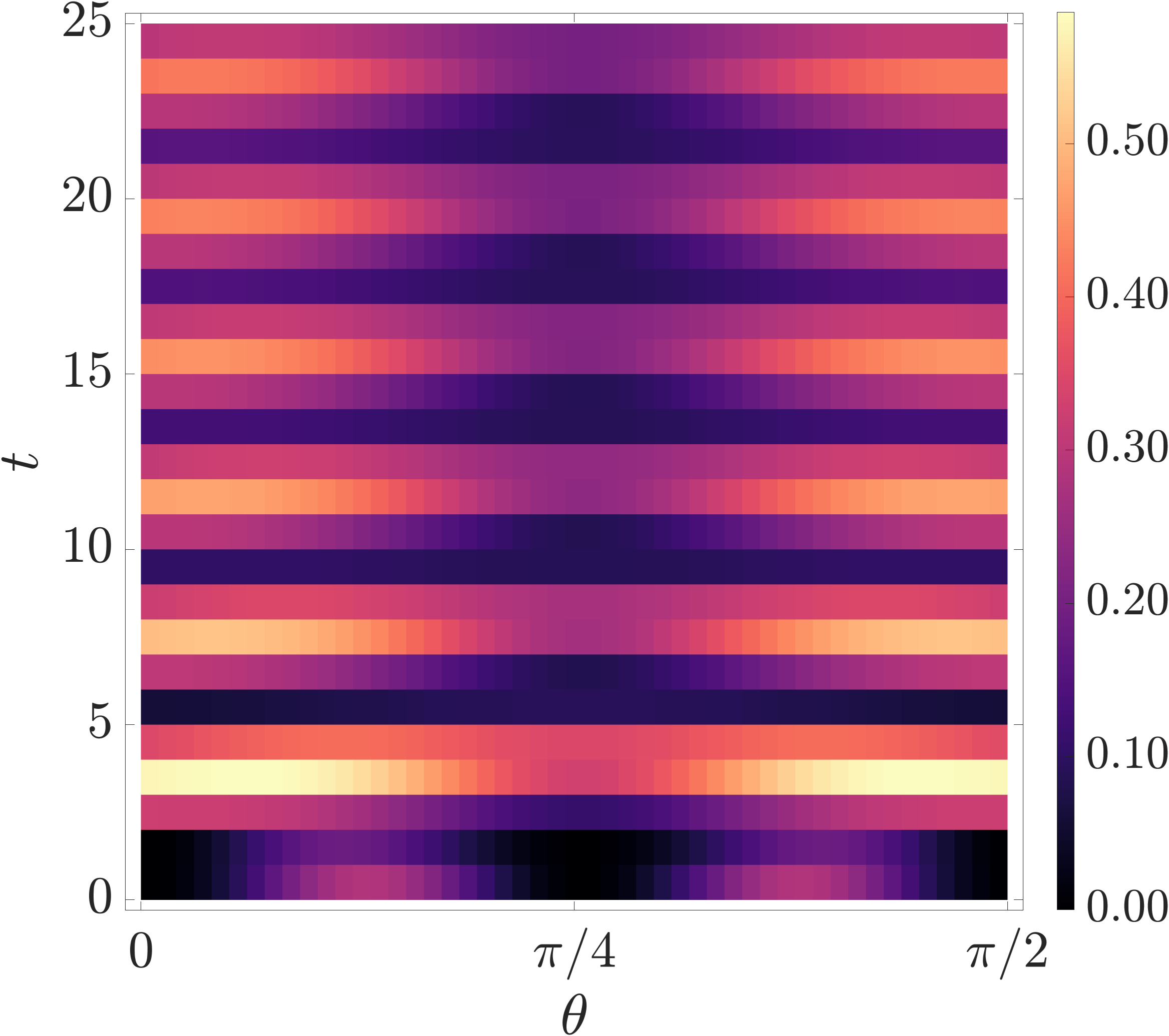}}
    \subfigure[]{\includegraphics[width=0.3\textwidth]{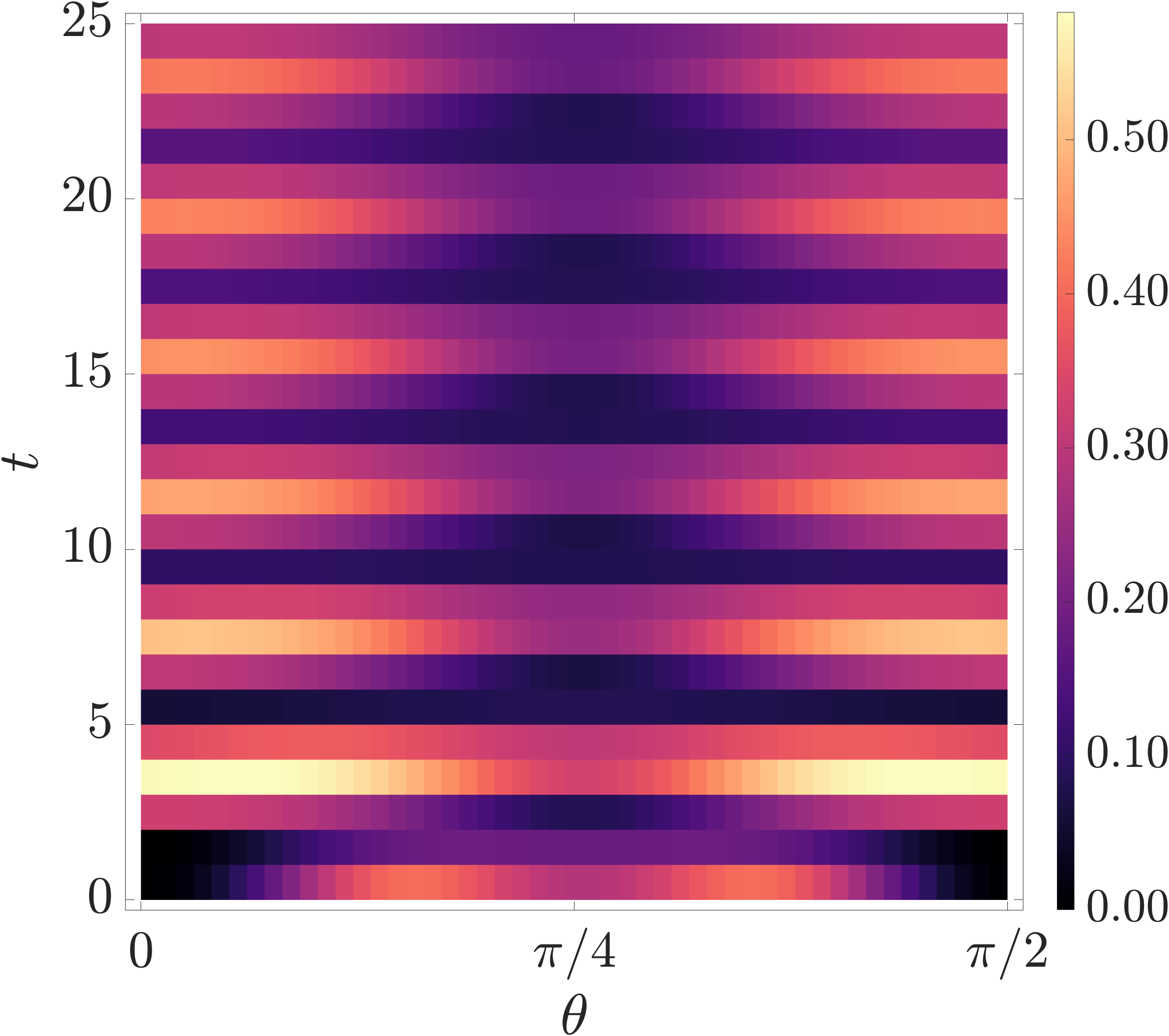}}
    \subfigure[]{\includegraphics[width=0.3\textwidth]{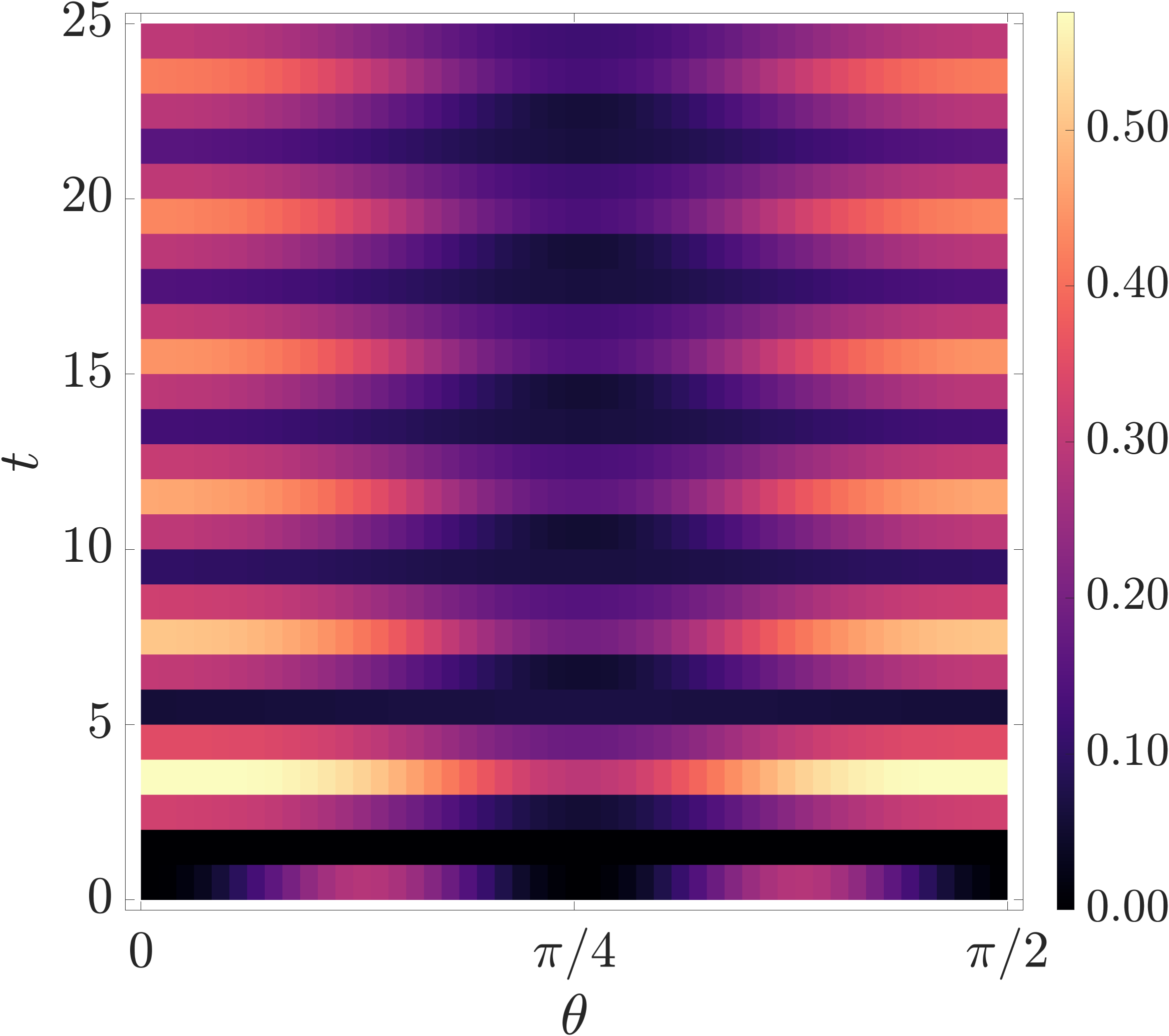}}
    \caption{The SRE as a function of time and the parameter $\theta$ in the initial state of the coins, $\ket{\chi} =  \cos(\theta/2) \ket{\uparrow \uparrow} + e^{i \phi} \sin(\theta/2) \ket{\downarrow \downarrow} $ for (a) $\phi = 0$, (b) $\phi = \pi/4$, (b) $\phi = \pi/2$. The dynamics involving two walkers retain the oscillatory pattern, as was the case with one walker, but with an additional symmetry. The system size is taken to be $101 \times 101$.}
    \label{fig:theta_t_two walkers}
\end{figure*}
\begin{figure}
    \centering
    \subfigure[]{\includegraphics[width=0.35\textwidth]{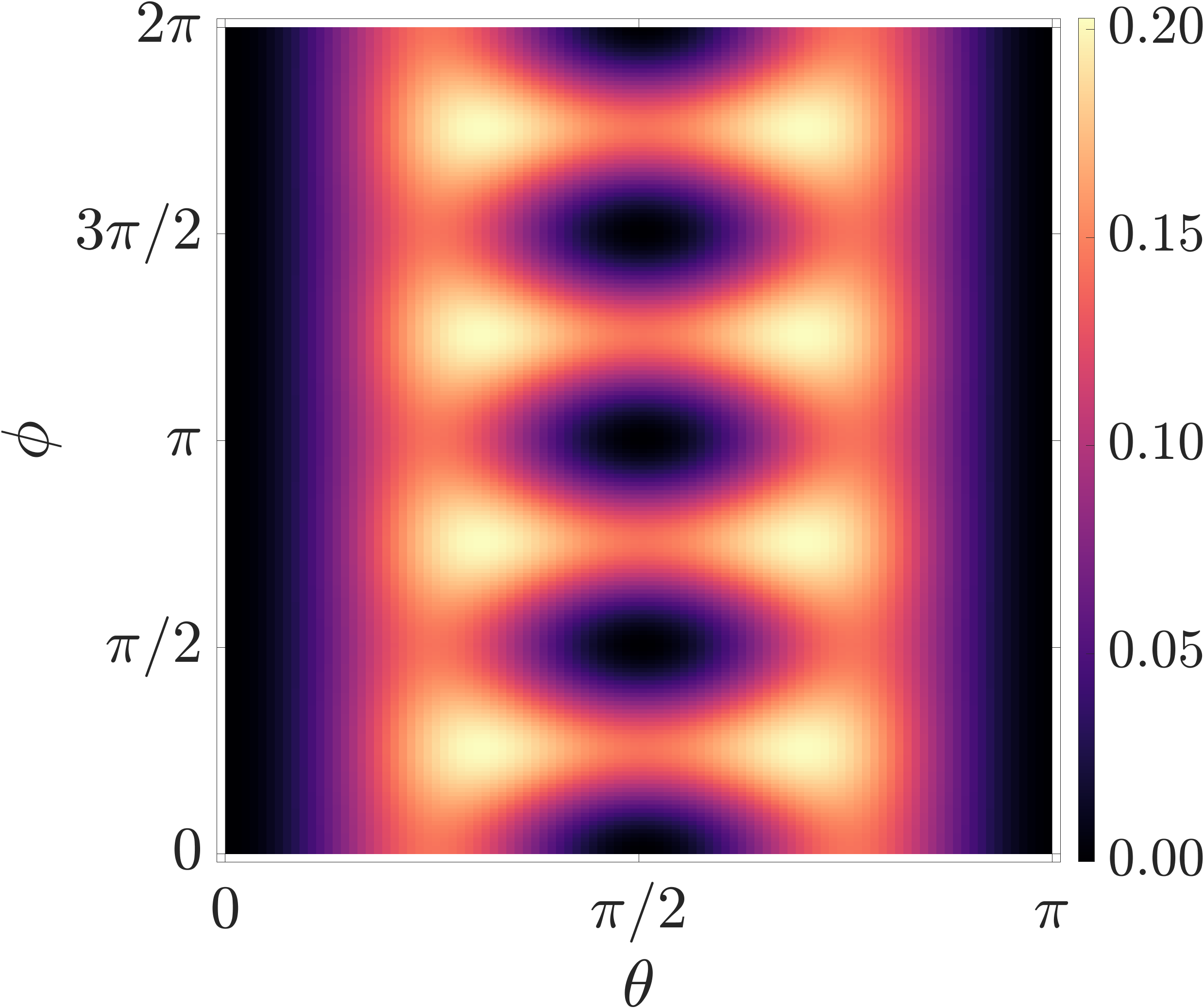}}
    \subfigure[]{\includegraphics[width=0.35\textwidth]{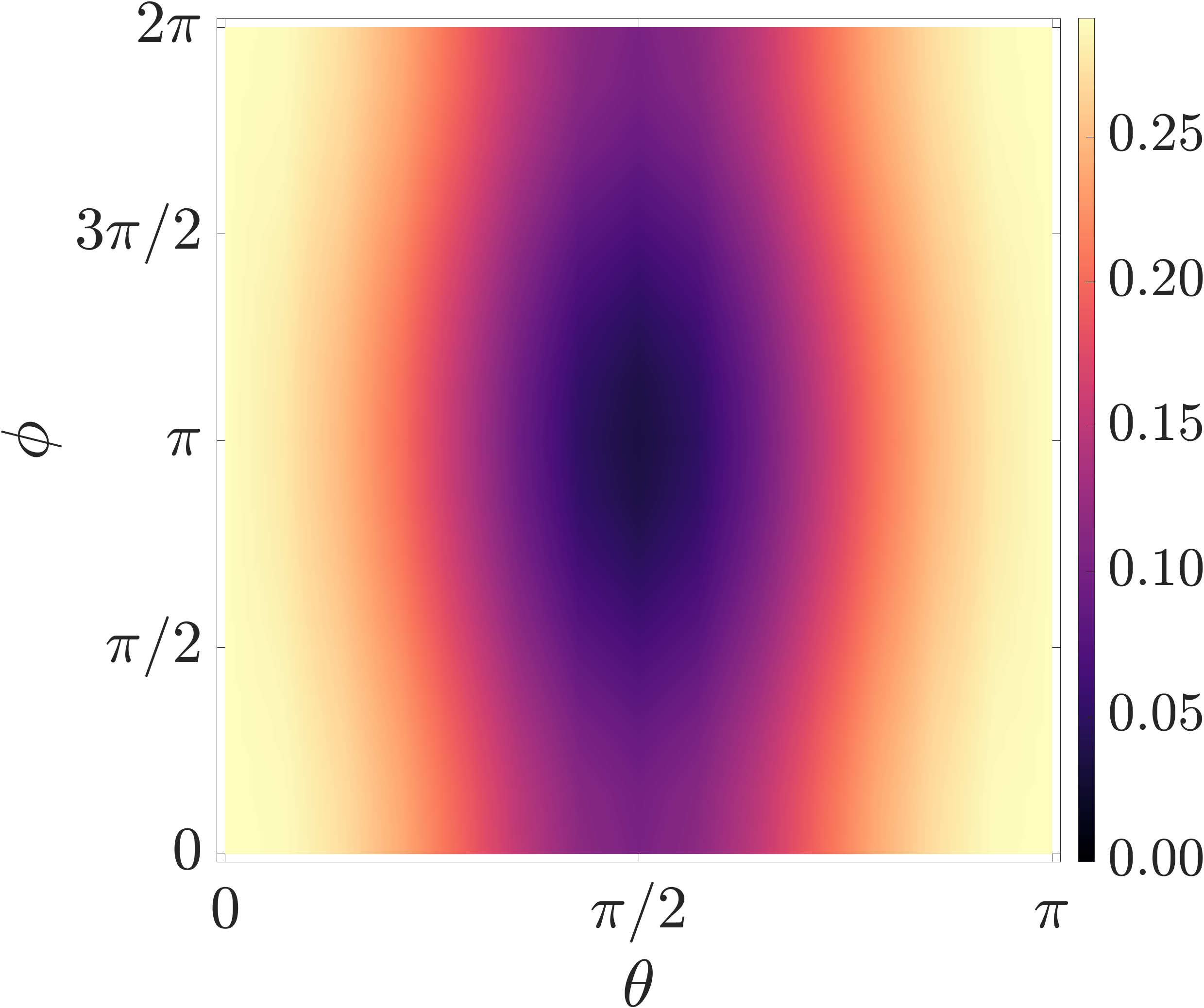}}
    \caption{The SRE for the different initial states of the coin given by Eq.~\eqref{eq:initialcoins} at (a) $t = 0$, (b) $t = 50$ time steps. We observe that the magic disperses in the system and has almost a correlation with the initial amount of magic in the system. The states with initially no magic evolve to a highly magical state. Here, the system size is taken to be $101 \times 101$.}
    \label{fig:t0_t50_twowalker}
\end{figure}
A natural extension would be to probe the quantum magic with two walkers that provide more freedom as the number of parameters in the initial state of the two coins increases. To keep the analysis simple, we consider a straightforward extension of the discrete-time quantum walk of one walker to get a quantum walk with two walkers. The structure of the composite Hilbert space is $\mathcal{H} = \mathcal{H}_{C_1} \otimes \mathcal{H}_{C_2} \otimes \mathcal{H}_{P_1} \otimes \mathcal{H}_{P_2}$, and correspondingly, the time evolution operator for a unit time step reads
\begin{equation}
    \bar{U} = (S_2 \cdot H ) (S_1 \cdot H) 
\end{equation}
where $S_1$ and $S_2$ are the shift operators for the first and second walker, respectively, and were introduced in the previous section. We use the Hadamard coin for both walkers. Note, we are skipping writing the identity explicitly to avoid clutter.

In this case, we initialize the walkers in localized states in position, and the coins to be in a superposition, which reads 
\begin{equation}
    \ket{\Psi(0)} = \ket{\chi} \otimes \ket{x_1 = 0} \otimes \ket{x_2 = 0}
    \label{eq:initialstate_two}
\end{equation}
with
\begin{equation}
    \label{eq:initialcoins}
    \ket{\chi} = \left( \cos(\theta/2) \ket{\uparrow \uparrow} + e^{i \phi} \sin(\theta/2) \ket{\downarrow \downarrow} \right)
\end{equation}
where we again use the Bloch sphere-like parameterization for the state of the two coins with $\theta \in [0, \pi]$ and $\phi \in [0, 2\pi]$. We choose this parametrization for the state of the two coins because it allows us to explore two extremes, a product state (for $\theta = 0$), and a maximally non-local entangled state (for $\theta = \pi/2$), for the case of pure states. 

Note that in this case, the full state space of a two-qubit system is huge (15-dimensional and 7-dimensional if we restrict ourselves to only pure states), which makes it impractical to perform an exhaustive scan over all possible configurations for the initial state of the two coins. Even numerically, a blind exploration of such a parameter space would be extremely demanding in terms of computational resources, while not necessarily providing deeper physical insight. To address this, we adopted a strategic and physically meaningful approach in selecting representative subspaces of initial states.

In Fig.~\ref{fig:fixedtheta_t_twowalkers}, we present the time evolution of SRE for four different composite initial coin states: $\ket{\chi} = \ket{\uparrow \uparrow}$, a Bell state $\ket{\chi} = \left( \ket{\uparrow \uparrow} + \ket{\downarrow \downarrow} \right)/\sqrt{2}$, a Bell state after the application of a magic gate $T$ to the first coin, and a two-qubit state with maximal magic~\cite{Zhewei2025}, $\ket{\chi} = \left( \ket{\uparrow \uparrow} + i \ket{\uparrow \downarrow} + i\ket{\downarrow \uparrow} + i\ket{\downarrow \downarrow} \right)/2$. The magic gate for a single qubit is given by
\begin{equation}
    T = \begin{pmatrix}
        1 & 0 \\
        0 & e^{i \pi/4}
    \end{pmatrix}.
\end{equation}
As expected, the application of the magic gate enhances the magic at time $t = 0$, confirming its role in injecting non-stabilizer components into the state. To our surprise, here, an entangled state, a magic-enhanced state, and the state with initially maximal magic show limited magic growth over time, while a simple product state produces significantly higher SRE, which is highly counterintuitive at this stage.

In Fig.~\ref{fig:theta_t_two walkers}, we generalize the observations of a single walker in Fig.~\ref{fig:theta_t} to the two-walker case. We are plotting the SRE as a function of time and $\theta$ for various fixed $\phi$ values in the allowed range from Eq.~\eqref{eq:initialstate_two}. For all phases investigated, the SRE again shows a sensitivity towards $\theta$, although the patterns are more structured and peaked compared to the single-walker scenario. In this case, we do not see a region where magic experiences a total death as the walk evolves. Also, we observe a periodic behavior with $\theta$ for all the values of the phase. The results indicate that magic generation in multi-walker systems is not as sensitive to the initial state of the system as in the case of a single walker. On the other hand, we see one similarity in this case, which was also present in the case of single-walker and at phase value $\phi = \pi/2$. The magic is zero after the first step of the quantum walk, irrespective of the initial amount of magic in the system.

We further compare the SRE for several initial coin states at two time points: $t = 0$ and $t = 50$ in Fig.~\ref{fig:t0_t50_twowalker}. Although all states begin with similar or zero SRE, their values diverge significantly over time. This divergence highlights the crucial role of dynamics in transforming initially classical or low-magic configurations into more complex states with a higher amount of magic. The results demonstrate that even states that are close to stabilizer form can evolve into highly magical states when subjected to quantum walk protocols. This showcases the potential of such systems in generating valuable computational resources.

\begin{figure}
    \centering
    \includegraphics[width=0.40\textwidth]{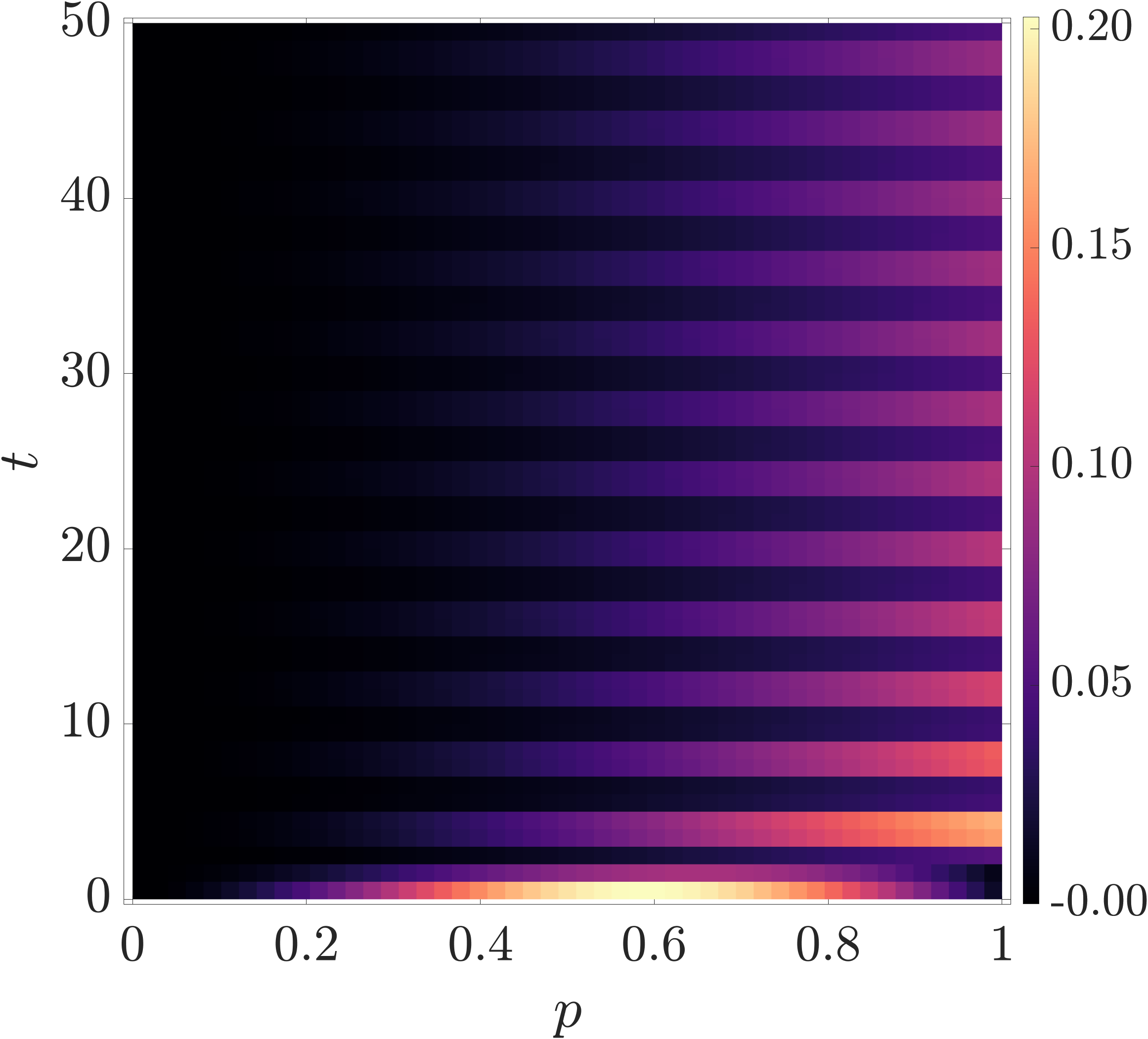}
    \caption{The SRE as a function of time for the case when two coins are initialized in a combined Werner state with parameter $p$. The Werner states comprise states with magic $\sim 0.20$; however, the dynamics of quantum walk suppress it as the system evolves. The system size is taken to be $101 \times 101$.}
    \label{fig:werner_t}
\end{figure}

Finally, we extend our analysis to include the mixed states for the choice of the initial state of the coins, for which we choose Werner states~\cite{Werner1989}, which again allows us to probe the two extremes, maximally mixed and maximally entangled states. Werner states are a family of mixed states that is given by a convex sum of a maximally entangled state and a maximally mixed state. It reads
\begin{equation}
    \label{eq:wernerstate}
    \rho(p) = p \dyad*{\Phi^+} + (1 - p) \dfrac{\mathds{1}_4}{4},
\end{equation}
where $\dyad*{\Phi^+} = \left( \ket{ \uparrow \uparrow} + \ket{\downarrow \downarrow} \right)/\sqrt{2}$ is a Bell state, $\mathds{1}/4$ is the maximally mixed state on the two-qubit Hilbert space, and $p \in [0,1]$ is a mixing parameter. For $p = 1$, the state is pure and maximally entangled, while for $p = 0$, it is fully mixed and contains no correlations, quantum or classical. In Fig.~\ref{fig:werner_t}, we again investigate the time evolution of the SRE for the reduced density matrix of the two coins states for different values of $p$ and observe oscillating behavior as in the previous cases. Surprisingly, we observe that magic can be generated dynamically even when the initial state has no magic ($p = 1$), whereas in other cases where the initial state has high magic ($p \approx 0.6$), the magic decays over time. This counterintuitive behavior highlights that the time evolution under the quantum walk is capable of redistributing and transforming quantum resources in a nontrivial way. It also suggests that initial magic is not always a reliable predictor of long-term behavior.

Note, for the sake of completeness, we also explored the initial states of the form,
\begin{equation}
    \ket{\chi} = \left( \cos(\theta/2) \ket{\uparrow \downarrow} + e^{i \phi} \sin(\theta/2) \ket{\downarrow \uparrow} \right)
\end{equation}
along with the initial state given in Eq.~\eqref{eq:initialcoins}, and we observed the same dynamics, qualitatively. 

\section{Conclusion \& Outlook}
\label{sec:conclusion}
In this work, we have shown that DTQWs serve as flexible generators and controllers of quantum magic. By evolving single and two-particle walks from various initial coin states, we found that quantum walk dynamics can both produce and reshape magic in nontrivial ways. For a single walker, large initial magic does not guarantee large asymptotic magic – the walk can amplify or diminish magic depending on interference effects.

Furthermore, our analytical results in the asymptotic time limit revealed a complementary trade-off between coin–position entanglement and quantum magic: maximal entanglement often coincides with minimal magic.  Although a complete mathematical inequality is still needed, our results provide an empirical foundation for future theoretical developments and can serve as a pivotal point. In the two-walkers case, even walks starting from stabilizer states become highly non-stabilizer due to coherent interference. We demonstrated that the generated magic in two-walker DTQWs can be tuned via the initial coin entanglement (for example, by varying Werner-state parameters), highlighting a high degree of control.

Importantly, our key findings are rooted in a theoretical model that directly connects to experimental implementation. Given that the Stabilizer R\'enyi Entropy relies solely on tomographic measurements of the reduced coin state and the Pauli measurements (which scale as $4^n$ with the number of qubits of $n$), the magic dynamics predicted in the one-dimensional case can be investigated in various experimental platforms. For example, photonic quantum walk systems using integrated waveguide arrays~\cite{Tang2018, Wang2020}, and atomic quantum walks in optical lattices~\cite{Preiss2015} have already demonstrated the precise control and measurement capabilities required for our protocol (specifically the number of time steps($\simeq50$), system size($\simeq100$)).
In Ref.~\cite{Wang2020}, the authors explored the robustness of entanglement as a marker for topological phases in quantum walks~\cite{Kitagawa2010, Asboth2012, Mittal2021} at asymptotic time scales. Furthermore, we have shown that this process is robust to realistic experimental imperfections, such as decoherence in the coin degree of freedom. Quantum magic persists across a broad range of noise strengths, suggesting that DTQWs can serve as reliable testbeds for studying dynamical resource theories in near-term quantum devices. This makes our predictions directly testable on existing quantum walk platforms. Our results bridge static resource theories and dynamical quantum walks, showing how non-Clifford resources naturally emerge in simple coherent evolutions.

Looking ahead, there are several natural extensions of this work. One direction is to consider more general classes of initial coin states—beyond the Bell-like superposition states and Werner-like mixtures studied here—in the two-walker setting, to examine how different forms of initial quantum correlations influence the subsequent dynamics of magic and entanglement. Another promising avenue is to explore more complex quantum walk architectures, such as higher-dimensional walks or systems with many walkers, to assess how magic generation scales with system size and dynamical complexity~\cite{Sticlet2025, Turkeshi2025a}. Recently, the magic generation has been proposed in a non-Hermitian setting~\cite{Chenu2025}, and it would be interesting to expand this analysis to quantum walks, as quantum walk systems have been instrumental in exploring non-Hermitian physics.

Finally, investigating the impact of noise and decoherence on magic generation—especially beyond minimal models—will be crucial for evaluating the practical viability of these ideas in near-term experiments and analog quantum simulators.

These results also raise an intriguing foundational question. In particular, from the complementary behavior of the quantum magic and quantum entanglement as shown in Fig.~\ref{fig:large_t}, one may ask whether there exists a conserved or reference-frame–invariant combination of quantum magic and entanglement, analogous to recent demonstrations of an invariant sum of entanglement and subsystem coherence under quantum reference frame transformations~\cite{Brukner2025}. A path to find a solution to this question could reveal deeper structural principles of dynamical quantum resource theory. In that spirit, discrete-time quantum walks may serve as a valuable platform for exploring these ideas, potentially revealing new principles of dynamical quantum resource theory. We need to see how far quantum walk systems can take us to understand these fundamental questions.

\section{Acknowledgments}
We would like to thank Fabrizio G. Oliviero for introducing us to the physics of quantum magic and for the useful discussions. V.M. expresses sincere gratitude to Navdeep Arya for his insightful discussions and valuable comments on the manuscript. This research has been supported by the MOST Young Scholar Fellowship (Grants No. 112-2636-M-007-008- No. 113-2636-M-007-002- and No. 114-2636-M-007 -001 -), National Center for Theoretical Sciences (Grants No. 113-2124-M-002-003-) from the Ministry of Science and Technology (MOST), Taiwan, Yushan Young Scholar Program (as Administrative Support Grant Fellow) from the Ministry of Education, Taiwan, and National Science and Technology Council (114-2811-M-008-075-MY2).


\begin{thebibliography}{71}%
	\makeatletter
	\providecommand \@ifxundefined [1]{%
		\@ifx{#1\undefined}
	}%
	\providecommand \@ifnum [1]{%
		\ifnum #1\expandafter \@firstoftwo
		\else \expandafter \@secondoftwo
		\fi
	}%
	\providecommand \@ifx [1]{%
		\ifx #1\expandafter \@firstoftwo
		\else \expandafter \@secondoftwo
		\fi
	}%
	\providecommand \natexlab [1]{#1}%
	\providecommand \enquote  [1]{``#1''}%
	\providecommand \bibnamefont  [1]{#1}%
	\providecommand \bibfnamefont [1]{#1}%
	\providecommand \citenamefont [1]{#1}%
	\providecommand \href@noop [0]{\@secondoftwo}%
	\providecommand \href [0]{\begingroup \@sanitize@url \@href}%
	\providecommand \@href[1]{\@@startlink{#1}\@@href}%
	\providecommand \@@href[1]{\endgroup#1\@@endlink}%
	\providecommand \@sanitize@url [0]{\catcode `\\12\catcode `\$12\catcode
		`\&12\catcode `\#12\catcode `\^12\catcode `\_12\catcode `\%12\relax}%
	\providecommand \@@startlink[1]{}%
	\providecommand \@@endlink[0]{}%
	\providecommand \url  [0]{\begingroup\@sanitize@url \@url }%
	\providecommand \@url [1]{\endgroup\@href {#1}{\urlprefix }}%
	\providecommand \urlprefix  [0]{URL }%
	\providecommand \Eprint [0]{\href }%
	\providecommand \doibase [0]{https://doi.org/}%
	\providecommand \selectlanguage [0]{\@gobble}%
	\providecommand \bibinfo  [0]{\@secondoftwo}%
	\providecommand \bibfield  [0]{\@secondoftwo}%
	\providecommand \translation [1]{[#1]}%
	\providecommand \BibitemOpen [0]{}%
	\providecommand \bibitemStop [0]{}%
	\providecommand \bibitemNoStop [0]{.\EOS\space}%
	\providecommand \EOS [0]{\spacefactor3000\relax}%
	\providecommand \BibitemShut  [1]{\csname bibitem#1\endcsname}%
	\let\auto@bib@innerbib\@empty
	\bibitem [{\citenamefont {Horodecki}\ \emph {et~al.}(2009)\citenamefont
		{Horodecki}, \citenamefont {Horodecki}, \citenamefont {Horodecki},\ and\
		\citenamefont {Horodecki}}]{Horodecki2009}%
	\BibitemOpen
	\bibfield  {author} {\bibinfo {author} {\bibfnamefont {R.}~\bibnamefont
			{Horodecki}}, \bibinfo {author} {\bibfnamefont {P.}~\bibnamefont
			{Horodecki}}, \bibinfo {author} {\bibfnamefont {M.}~\bibnamefont
			{Horodecki}},\ and\ \bibinfo {author} {\bibfnamefont {K.}~\bibnamefont
			{Horodecki}},\ }\bibfield  {title} {\bibinfo {title} {Quantum entanglement},\
	}\href {https://doi.org/10.1103/RevModPhys.81.865} {\bibfield  {journal}
		{\bibinfo  {journal} {Rev. Mod. Phys.}\ }\textbf {\bibinfo {volume} {81}},\
		\bibinfo {pages} {865} (\bibinfo {year} {2009})}\BibitemShut {NoStop}%
	\bibitem [{\citenamefont {Brunner}\ \emph {et~al.}(2014)\citenamefont
		{Brunner}, \citenamefont {Cavalcanti}, \citenamefont {Pironio}, \citenamefont
		{Scarani},\ and\ \citenamefont {Wehner}}]{Brunner2014}%
	\BibitemOpen
	\bibfield  {author} {\bibinfo {author} {\bibfnamefont {N.}~\bibnamefont
			{Brunner}}, \bibinfo {author} {\bibfnamefont {D.}~\bibnamefont {Cavalcanti}},
		\bibinfo {author} {\bibfnamefont {S.}~\bibnamefont {Pironio}}, \bibinfo
		{author} {\bibfnamefont {V.}~\bibnamefont {Scarani}},\ and\ \bibinfo {author}
		{\bibfnamefont {S.}~\bibnamefont {Wehner}},\ }\bibfield  {title} {\bibinfo
		{title} {Bell nonlocality},\ }\href
	{https://doi.org/10.1103/RevModPhys.86.419} {\bibfield  {journal} {\bibinfo
			{journal} {Rev. Mod. Phys.}\ }\textbf {\bibinfo {volume} {86}},\ \bibinfo
		{pages} {419} (\bibinfo {year} {2014})}\BibitemShut {NoStop}%
	\bibitem [{\citenamefont {Streltsov}\ \emph {et~al.}(2017)\citenamefont
		{Streltsov}, \citenamefont {Adesso},\ and\ \citenamefont
		{Plenio}}]{Adesso2017}%
	\BibitemOpen
	\bibfield  {author} {\bibinfo {author} {\bibfnamefont {A.}~\bibnamefont
			{Streltsov}}, \bibinfo {author} {\bibfnamefont {G.}~\bibnamefont {Adesso}},\
		and\ \bibinfo {author} {\bibfnamefont {M.~B.}\ \bibnamefont {Plenio}},\
	}\bibfield  {title} {\bibinfo {title} {Colloquium: Quantum coherence as a
			resource},\ }\href {https://doi.org/10.1103/RevModPhys.89.041003} {\bibfield
		{journal} {\bibinfo  {journal} {Rev. Mod. Phys.}\ }\textbf {\bibinfo {volume}
			{89}},\ \bibinfo {pages} {041003} (\bibinfo {year} {2017})}\BibitemShut
	{NoStop}%
	\bibitem [{\citenamefont {{Gottesman}}(1997)}]{Gottesman1997}%
	\BibitemOpen
	\bibfield  {author} {\bibinfo {author} {\bibfnamefont {D.}~\bibnamefont
			{{Gottesman}}},\ }\emph {\bibinfo {title} {{Stabilizer codes and quantum
				error correction}}},\ \href@noop {} {Ph.D. thesis},\ \bibinfo  {school}
	{California Institute of Technology} (\bibinfo {year} {1997})\BibitemShut
	{NoStop}%
	\bibitem [{\citenamefont {{Gottesman}}(1998)}]{Gottesman1998}%
	\BibitemOpen
	\bibfield  {author} {\bibinfo {author} {\bibfnamefont {D.}~\bibnamefont
			{{Gottesman}}},\ }\bibfield  {title} {\bibinfo {title} {{The Heisenberg
				Representation of Quantum Computers}},\ }\href
	{https://doi.org/10.48550/arXiv.quant-ph/9807006} {\bibfield  {journal}
		{\bibinfo  {journal} {arXiv e-prints}\ ,\ \bibinfo {eid} {quant-ph/9807006}}
		(\bibinfo {year} {1998})}\BibitemShut {NoStop}%
	\bibitem [{\citenamefont {Aaronson}\ and\ \citenamefont
		{Gottesman}(2004)}]{Aaronson2004}%
	\BibitemOpen
	\bibfield  {author} {\bibinfo {author} {\bibfnamefont {S.}~\bibnamefont
			{Aaronson}}\ and\ \bibinfo {author} {\bibfnamefont {D.}~\bibnamefont
			{Gottesman}},\ }\bibfield  {title} {\bibinfo {title} {Improved simulation of
			stabilizer circuits},\ }\href {https://doi.org/10.1103/PhysRevA.70.052328}
	{\bibfield  {journal} {\bibinfo  {journal} {Phys. Rev. A}\ }\textbf {\bibinfo
			{volume} {70}},\ \bibinfo {pages} {052328} (\bibinfo {year}
		{2004})}\BibitemShut {NoStop}%
	\bibitem [{\citenamefont {Bravyi}\ and\ \citenamefont
		{Kitaev}(2005)}]{Kitaev2005}%
	\BibitemOpen
	\bibfield  {author} {\bibinfo {author} {\bibfnamefont {S.}~\bibnamefont
			{Bravyi}}\ and\ \bibinfo {author} {\bibfnamefont {A.}~\bibnamefont
			{Kitaev}},\ }\bibfield  {title} {\bibinfo {title} {Universal quantum
			computation with ideal clifford gates and noisy ancillas},\ }\href
	{https://doi.org/10.1103/PhysRevA.71.022316} {\bibfield  {journal} {\bibinfo
			{journal} {Phys. Rev. A}\ }\textbf {\bibinfo {volume} {71}},\ \bibinfo
		{pages} {022316} (\bibinfo {year} {2005})}\BibitemShut {NoStop}%
	\bibitem [{\citenamefont {Veitch}\ \emph {et~al.}(2014)\citenamefont {Veitch},
		\citenamefont {Hamed~Mousavian}, \citenamefont {Gottesman},\ and\
		\citenamefont {Emerson}}]{Veitch2014}%
	\BibitemOpen
	\bibfield  {author} {\bibinfo {author} {\bibfnamefont {V.}~\bibnamefont
			{Veitch}}, \bibinfo {author} {\bibfnamefont {S.~A.}\ \bibnamefont
			{Hamed~Mousavian}}, \bibinfo {author} {\bibfnamefont {D.}~\bibnamefont
			{Gottesman}},\ and\ \bibinfo {author} {\bibfnamefont {J.}~\bibnamefont
			{Emerson}},\ }\bibfield  {title} {\bibinfo {title} {The resource theory of
			stabilizer quantum computation},\ }\href
	{https://doi.org/10.1088/1367-2630/16/1/013009} {\bibfield  {journal}
		{\bibinfo  {journal} {New Journal of Physics}\ }\textbf {\bibinfo {volume}
			{16}},\ \bibinfo {pages} {013009} (\bibinfo {year} {2014})}\BibitemShut
	{NoStop}%
	\bibitem [{\citenamefont {Bravyi}\ \emph {et~al.}(2016)\citenamefont {Bravyi},
		\citenamefont {Smith},\ and\ \citenamefont {Smolin}}]{Kitaev2016}%
	\BibitemOpen
	\bibfield  {author} {\bibinfo {author} {\bibfnamefont {S.}~\bibnamefont
			{Bravyi}}, \bibinfo {author} {\bibfnamefont {G.}~\bibnamefont {Smith}},\ and\
		\bibinfo {author} {\bibfnamefont {J.~A.}\ \bibnamefont {Smolin}},\ }\bibfield
	{title} {\bibinfo {title} {Trading classical and quantum computational
			resources},\ }\href {https://doi.org/10.1103/PhysRevX.6.021043} {\bibfield
		{journal} {\bibinfo  {journal} {Phys. Rev. X}\ }\textbf {\bibinfo {volume}
			{6}},\ \bibinfo {pages} {021043} (\bibinfo {year} {2016})}\BibitemShut
	{NoStop}%
	\bibitem [{\citenamefont {Chitambar}\ and\ \citenamefont
		{Gour}(2019)}]{Gour2019}%
	\BibitemOpen
	\bibfield  {author} {\bibinfo {author} {\bibfnamefont {E.}~\bibnamefont
			{Chitambar}}\ and\ \bibinfo {author} {\bibfnamefont {G.}~\bibnamefont
			{Gour}},\ }\bibfield  {title} {\bibinfo {title} {Quantum resource theories},\
	}\href {https://doi.org/10.1103/RevModPhys.91.025001} {\bibfield  {journal}
		{\bibinfo  {journal} {Rev. Mod. Phys.}\ }\textbf {\bibinfo {volume} {91}},\
		\bibinfo {pages} {025001} (\bibinfo {year} {2019})}\BibitemShut {NoStop}%
	\bibitem [{\citenamefont {Liu}\ and\ \citenamefont
		{Winter}(2022)}]{Winter2022}%
	\BibitemOpen
	\bibfield  {author} {\bibinfo {author} {\bibfnamefont {Z.-W.}\ \bibnamefont
			{Liu}}\ and\ \bibinfo {author} {\bibfnamefont {A.}~\bibnamefont {Winter}},\
	}\bibfield  {title} {\bibinfo {title} {Many-body quantum magic},\ }\href
	{https://doi.org/10.1103/PRXQuantum.3.020333} {\bibfield  {journal} {\bibinfo
			{journal} {PRX Quantum}\ }\textbf {\bibinfo {volume} {3}},\ \bibinfo {pages}
		{020333} (\bibinfo {year} {2022})}\BibitemShut {NoStop}%
	\bibitem [{\citenamefont {Dai}\ \emph {et~al.}(2022)\citenamefont {Dai},
		\citenamefont {Fu},\ and\ \citenamefont {Luo}}]{Dai2022}%
	\BibitemOpen
	\bibfield  {author} {\bibinfo {author} {\bibfnamefont {H.}~\bibnamefont
			{Dai}}, \bibinfo {author} {\bibfnamefont {S.}~\bibnamefont {Fu}},\ and\
		\bibinfo {author} {\bibfnamefont {S.}~\bibnamefont {Luo}},\ }\bibfield
	{title} {\bibinfo {title} {Detecting magic states via characteristic
			functions},\ }\href {https://doi.org/10.1007/s10773-022-05027-8} {\bibfield
		{journal} {\bibinfo  {journal} {International Journal of Theoretical
				Physics}\ }\textbf {\bibinfo {volume} {61}},\ \bibinfo {pages} {35} (\bibinfo
		{year} {2022})}\BibitemShut {NoStop}%
	\bibitem [{\citenamefont {Tirrito}\ \emph {et~al.}(2024)\citenamefont
		{Tirrito}, \citenamefont {Tarabunga}, \citenamefont {Lami}, \citenamefont
		{Chanda}, \citenamefont {Leone}, \citenamefont {Oliviero}, \citenamefont
		{Dalmonte}, \citenamefont {Collura},\ and\ \citenamefont
		{Hamma}}]{Marcello2024}%
	\BibitemOpen
	\bibfield  {author} {\bibinfo {author} {\bibfnamefont {E.}~\bibnamefont
			{Tirrito}}, \bibinfo {author} {\bibfnamefont {P.~S.}\ \bibnamefont
			{Tarabunga}}, \bibinfo {author} {\bibfnamefont {G.}~\bibnamefont {Lami}},
		\bibinfo {author} {\bibfnamefont {T.}~\bibnamefont {Chanda}}, \bibinfo
		{author} {\bibfnamefont {L.}~\bibnamefont {Leone}}, \bibinfo {author}
		{\bibfnamefont {S.~F.~E.}\ \bibnamefont {Oliviero}}, \bibinfo {author}
		{\bibfnamefont {M.}~\bibnamefont {Dalmonte}}, \bibinfo {author}
		{\bibfnamefont {M.}~\bibnamefont {Collura}},\ and\ \bibinfo {author}
		{\bibfnamefont {A.}~\bibnamefont {Hamma}},\ }\bibfield  {title} {\bibinfo
		{title} {Quantifying nonstabilizerness through entanglement spectrum
			flatness},\ }\href {https://doi.org/10.1103/PhysRevA.109.L040401} {\bibfield
		{journal} {\bibinfo  {journal} {Phys. Rev. A}\ }\textbf {\bibinfo {volume}
			{109}},\ \bibinfo {pages} {L040401} (\bibinfo {year} {2024})}\BibitemShut
	{NoStop}%
	\bibitem [{\citenamefont {De~Nicola}\ \emph {et~al.}(2014)\citenamefont
		{De~Nicola}, \citenamefont {Sansoni}, \citenamefont {Crespi}, \citenamefont
		{Ramponi}, \citenamefont {Osellame}, \citenamefont {Giovannetti},
		\citenamefont {Fazio}, \citenamefont {Mataloni},\ and\ \citenamefont
		{Sciarrino}}]{Nicola2014}%
	\BibitemOpen
	\bibfield  {author} {\bibinfo {author} {\bibfnamefont {F.}~\bibnamefont
			{De~Nicola}}, \bibinfo {author} {\bibfnamefont {L.}~\bibnamefont {Sansoni}},
		\bibinfo {author} {\bibfnamefont {A.}~\bibnamefont {Crespi}}, \bibinfo
		{author} {\bibfnamefont {R.}~\bibnamefont {Ramponi}}, \bibinfo {author}
		{\bibfnamefont {R.}~\bibnamefont {Osellame}}, \bibinfo {author}
		{\bibfnamefont {V.}~\bibnamefont {Giovannetti}}, \bibinfo {author}
		{\bibfnamefont {R.}~\bibnamefont {Fazio}}, \bibinfo {author} {\bibfnamefont
			{P.}~\bibnamefont {Mataloni}},\ and\ \bibinfo {author} {\bibfnamefont
			{F.}~\bibnamefont {Sciarrino}},\ }\bibfield  {title} {\bibinfo {title}
		{Quantum simulation of {b}osonic-{f}ermionic noninteracting particles in
			disordered systems via a quantum walk},\ }\href
	{https://doi.org/10.1103/PhysRevA.89.032322} {\bibfield  {journal} {\bibinfo
			{journal} {Phys. Rev. A}\ }\textbf {\bibinfo {volume} {89}},\ \bibinfo
		{pages} {032322} (\bibinfo {year} {2014})}\BibitemShut {NoStop}%
	\bibitem [{\citenamefont {{Zhang}}\ and\ \citenamefont
		{{Gu}}(2024)}]{Zhang2024}%
	\BibitemOpen
	\bibfield  {author} {\bibinfo {author} {\bibfnamefont {Y.}~\bibnamefont
			{{Zhang}}}\ and\ \bibinfo {author} {\bibfnamefont {Y.}~\bibnamefont {{Gu}}},\
	}\bibfield  {title} {\bibinfo {title} {{Quantum magic dynamics in random
				circuits}},\ }\href {https://doi.org/10.48550/arXiv.2410.21128} {\bibfield
		{journal} {\bibinfo  {journal} {arXiv e-prints}\ ,\ \bibinfo {eid}
			{arXiv:2410.21128}} (\bibinfo {year} {2024})}\BibitemShut {NoStop}%
	\bibitem [{\citenamefont {Turkeshi}\ \emph
		{et~al.}(2025{\natexlab{a}})\citenamefont {Turkeshi}, \citenamefont
		{Dymarsky},\ and\ \citenamefont {Sierant}}]{Piotr2025}%
	\BibitemOpen
	\bibfield  {author} {\bibinfo {author} {\bibfnamefont {X.}~\bibnamefont
			{Turkeshi}}, \bibinfo {author} {\bibfnamefont {A.}~\bibnamefont {Dymarsky}},\
		and\ \bibinfo {author} {\bibfnamefont {P.}~\bibnamefont {Sierant}},\
	}\bibfield  {title} {\bibinfo {title} {Pauli spectrum and nonstabilizerness
			of typical quantum many-body states},\ }\href
	{https://doi.org/10.1103/PhysRevB.111.054301} {\bibfield  {journal} {\bibinfo
			{journal} {Phys. Rev. B}\ }\textbf {\bibinfo {volume} {111}},\ \bibinfo
		{pages} {054301} (\bibinfo {year} {2025}{\natexlab{a}})}\BibitemShut
	{NoStop}%
	\bibitem [{\citenamefont {Turkeshi}\ \emph
		{et~al.}(2025{\natexlab{b}})\citenamefont {Turkeshi}, \citenamefont
		{Tirrito},\ and\ \citenamefont {Sierant}}]{Turkeshi2025}%
	\BibitemOpen
	\bibfield  {author} {\bibinfo {author} {\bibfnamefont {X.}~\bibnamefont
			{Turkeshi}}, \bibinfo {author} {\bibfnamefont {E.}~\bibnamefont {Tirrito}},\
		and\ \bibinfo {author} {\bibfnamefont {P.}~\bibnamefont {Sierant}},\
	}\bibfield  {title} {\bibinfo {title} {Magic spreading in random quantum
			circuits},\ }\href {https://doi.org/10.1038/s41467-025-57704-x} {\bibfield
		{journal} {\bibinfo  {journal} {Nature Communications}\ }\textbf {\bibinfo
			{volume} {16}},\ \bibinfo {pages} {2575} (\bibinfo {year}
		{2025}{\natexlab{b}})}\BibitemShut {NoStop}%
	\bibitem [{\citenamefont {Fux}\ \emph {et~al.}(2024)\citenamefont {Fux},
		\citenamefont {Tirrito}, \citenamefont {Dalmonte},\ and\ \citenamefont
		{Fazio}}]{Rosario2024}%
	\BibitemOpen
	\bibfield  {author} {\bibinfo {author} {\bibfnamefont {G.~E.}\ \bibnamefont
			{Fux}}, \bibinfo {author} {\bibfnamefont {E.}~\bibnamefont {Tirrito}},
		\bibinfo {author} {\bibfnamefont {M.}~\bibnamefont {Dalmonte}},\ and\
		\bibinfo {author} {\bibfnamefont {R.}~\bibnamefont {Fazio}},\ }\bibfield
	{title} {\bibinfo {title} {Entanglement -- nonstabilizerness separation in
			hybrid quantum circuits},\ }\href
	{https://doi.org/10.1103/PhysRevResearch.6.L042030} {\bibfield  {journal}
		{\bibinfo  {journal} {Phys. Rev. Res.}\ }\textbf {\bibinfo {volume} {6}},\
		\bibinfo {pages} {L042030} (\bibinfo {year} {2024})}\BibitemShut {NoStop}%
	\bibitem [{\citenamefont {Bejan}\ \emph {et~al.}(2024)\citenamefont {Bejan},
		\citenamefont {McLauchlan},\ and\ \citenamefont {B\'eri}}]{Beri2024}%
	\BibitemOpen
	\bibfield  {author} {\bibinfo {author} {\bibfnamefont {M.}~\bibnamefont
			{Bejan}}, \bibinfo {author} {\bibfnamefont {C.}~\bibnamefont {McLauchlan}},\
		and\ \bibinfo {author} {\bibfnamefont {B.}~\bibnamefont {B\'eri}},\
	}\bibfield  {title} {\bibinfo {title} {{Dynamical Magic Transitions in
				Monitored Clifford+$T$ Circuits}},\ }\href
	{https://doi.org/10.1103/PRXQuantum.5.030332} {\bibfield  {journal} {\bibinfo
			{journal} {PRX Quantum}\ }\textbf {\bibinfo {volume} {5}},\ \bibinfo {pages}
		{030332} (\bibinfo {year} {2024})}\BibitemShut {NoStop}%
	\bibitem [{\citenamefont {Niroula}\ \emph {et~al.}(2024)\citenamefont
		{Niroula}, \citenamefont {White}, \citenamefont {Wang}, \citenamefont
		{Johri}, \citenamefont {Zhu}, \citenamefont {Monroe}, \citenamefont {Noel},\
		and\ \citenamefont {Gullans}}]{Niroula2024}%
	\BibitemOpen
	\bibfield  {author} {\bibinfo {author} {\bibfnamefont {P.}~\bibnamefont
			{Niroula}}, \bibinfo {author} {\bibfnamefont {C.~D.}\ \bibnamefont {White}},
		\bibinfo {author} {\bibfnamefont {Q.}~\bibnamefont {Wang}}, \bibinfo {author}
		{\bibfnamefont {S.}~\bibnamefont {Johri}}, \bibinfo {author} {\bibfnamefont
			{D.}~\bibnamefont {Zhu}}, \bibinfo {author} {\bibfnamefont {C.}~\bibnamefont
			{Monroe}}, \bibinfo {author} {\bibfnamefont {C.}~\bibnamefont {Noel}},\ and\
		\bibinfo {author} {\bibfnamefont {M.~J.}\ \bibnamefont {Gullans}},\
	}\bibfield  {title} {\bibinfo {title} {Phase transition in magic with random
			quantum circuits},\ }\href {https://doi.org/10.1038/s41567-024-02637-3}
	{\bibfield  {journal} {\bibinfo  {journal} {Nature Physics}\ }\textbf
		{\bibinfo {volume} {20}},\ \bibinfo {pages} {1786} (\bibinfo {year}
		{2024})}\BibitemShut {NoStop}%
	\bibitem [{\citenamefont {Turkeshi}\ and\ \citenamefont
		{Sierant}(2024)}]{Turkeshi2024}%
	\BibitemOpen
	\bibfield  {author} {\bibinfo {author} {\bibfnamefont {X.}~\bibnamefont
			{Turkeshi}}\ and\ \bibinfo {author} {\bibfnamefont {P.}~\bibnamefont
			{Sierant}},\ }\bibfield  {title} {\bibinfo {title} {Error-resilience phase
			transitions in encoding-decoding quantum circuits},\ }\href
	{https://doi.org/10.1103/PhysRevLett.132.140401} {\bibfield  {journal}
		{\bibinfo  {journal} {Phys. Rev. Lett.}\ }\textbf {\bibinfo {volume} {132}},\
		\bibinfo {pages} {140401} (\bibinfo {year} {2024})}\BibitemShut {NoStop}%
	\bibitem [{\citenamefont {Aharonov}\ \emph {et~al.}(1993)\citenamefont
		{Aharonov}, \citenamefont {Davidovich},\ and\ \citenamefont
		{Zagury}}]{Aharonov1993}%
	\BibitemOpen
	\bibfield  {author} {\bibinfo {author} {\bibfnamefont {Y.}~\bibnamefont
			{Aharonov}}, \bibinfo {author} {\bibfnamefont {L.}~\bibnamefont
			{Davidovich}},\ and\ \bibinfo {author} {\bibfnamefont {N.}~\bibnamefont
			{Zagury}},\ }\bibfield  {title} {\bibinfo {title} {Quantum random walks},\
	}\href {https://doi.org/10.1103/PhysRevA.48.1687} {\bibfield  {journal}
		{\bibinfo  {journal} {Phys. Rev. A}\ }\textbf {\bibinfo {volume} {48}},\
		\bibinfo {pages} {1687} (\bibinfo {year} {1993})}\BibitemShut {NoStop}%
	\bibitem [{\citenamefont {Kempe}(2003)}]{Kempe2003}%
	\BibitemOpen
	\bibfield  {author} {\bibinfo {author} {\bibfnamefont {J.}~\bibnamefont
			{Kempe}},\ }\bibfield  {title} {\bibinfo {title} {Quantum random walks: An
			introductory overview},\ }\href
	{https://doi.org/10.1080/00107151031000110776} {\bibfield  {journal}
		{\bibinfo  {journal} {Contemp. Phys.}\ }\textbf {\bibinfo {volume} {44}},\
		\bibinfo {pages} {307} (\bibinfo {year} {2003})}\BibitemShut {NoStop}%
	\bibitem [{\citenamefont {Regensburger}\ \emph {et~al.}(2012)\citenamefont
		{Regensburger}, \citenamefont {Bersch}, \citenamefont {Miri}, \citenamefont
		{Onishchukov}, \citenamefont {Christodoulides},\ and\ \citenamefont
		{Peschel}}]{Regensburger2012}%
	\BibitemOpen
	\bibfield  {author} {\bibinfo {author} {\bibfnamefont {A.}~\bibnamefont
			{Regensburger}}, \bibinfo {author} {\bibfnamefont {C.}~\bibnamefont
			{Bersch}}, \bibinfo {author} {\bibfnamefont {M.-A.}\ \bibnamefont {Miri}},
		\bibinfo {author} {\bibfnamefont {G.}~\bibnamefont {Onishchukov}}, \bibinfo
		{author} {\bibfnamefont {D.~N.}\ \bibnamefont {Christodoulides}},\ and\
		\bibinfo {author} {\bibfnamefont {U.}~\bibnamefont {Peschel}},\ }\bibfield
	{title} {\bibinfo {title} {Parity--time synthetic photonic lattices},\ }\href
	{https://doi.org/10.1038/nature11298} {\bibfield  {journal} {\bibinfo
			{journal} {Nature (London)}\ }\textbf {\bibinfo {volume} {488}},\ \bibinfo
		{pages} {167} (\bibinfo {year} {2012})}\BibitemShut {NoStop}%
	\bibitem [{\citenamefont {Schreiber}\ \emph {et~al.}(2010)\citenamefont
		{Schreiber}, \citenamefont {Cassemiro}, \citenamefont
		{Poto\ifmmode~\check{c}\else \v{c}\fi{}ek}, \citenamefont {G\'abris},
		\citenamefont {Mosley}, \citenamefont {Andersson}, \citenamefont {Jex},\ and\
		\citenamefont {Silberhorn}}]{Schreiber2010}%
	\BibitemOpen
	\bibfield  {author} {\bibinfo {author} {\bibfnamefont {A.}~\bibnamefont
			{Schreiber}}, \bibinfo {author} {\bibfnamefont {K.~N.}\ \bibnamefont
			{Cassemiro}}, \bibinfo {author} {\bibfnamefont {V.}~\bibnamefont
			{Poto\ifmmode~\check{c}\else \v{c}\fi{}ek}}, \bibinfo {author} {\bibfnamefont
			{A.}~\bibnamefont {G\'abris}}, \bibinfo {author} {\bibfnamefont {P.~J.}\
			\bibnamefont {Mosley}}, \bibinfo {author} {\bibfnamefont {E.}~\bibnamefont
			{Andersson}}, \bibinfo {author} {\bibfnamefont {I.}~\bibnamefont {Jex}},\
		and\ \bibinfo {author} {\bibfnamefont {C.}~\bibnamefont {Silberhorn}},\
	}\bibfield  {title} {\bibinfo {title} {Photons walking the line: A quantum
			walk with adjustable coin operations},\ }\href
	{https://doi.org/10.1103/PhysRevLett.104.050502} {\bibfield  {journal}
		{\bibinfo  {journal} {Phys. Rev. Lett.}\ }\textbf {\bibinfo {volume} {104}},\
		\bibinfo {pages} {050502} (\bibinfo {year} {2010})}\BibitemShut {NoStop}%
	\bibitem [{\citenamefont {Schreiber}\ \emph {et~al.}(2012)\citenamefont
		{Schreiber}, \citenamefont {G{\'a}bris}, \citenamefont {Rohde}, \citenamefont
		{Laiho}, \citenamefont {{\v S}tefa{\v n}{\'a}k}, \citenamefont {Poto{\v
				c}ek}, \citenamefont {Hamilton}, \citenamefont {Jex},\ and\ \citenamefont
		{Silberhorn}}]{Schreiber2012}%
	\BibitemOpen
	\bibfield  {author} {\bibinfo {author} {\bibfnamefont {A.}~\bibnamefont
			{Schreiber}}, \bibinfo {author} {\bibfnamefont {A.}~\bibnamefont
			{G{\'a}bris}}, \bibinfo {author} {\bibfnamefont {P.~P.}\ \bibnamefont
			{Rohde}}, \bibinfo {author} {\bibfnamefont {K.}~\bibnamefont {Laiho}},
		\bibinfo {author} {\bibfnamefont {M.}~\bibnamefont {{\v S}tefa{\v n}{\'a}k}},
		\bibinfo {author} {\bibfnamefont {V.}~\bibnamefont {Poto{\v c}ek}}, \bibinfo
		{author} {\bibfnamefont {C.}~\bibnamefont {Hamilton}}, \bibinfo {author}
		{\bibfnamefont {I.}~\bibnamefont {Jex}},\ and\ \bibinfo {author}
		{\bibfnamefont {C.}~\bibnamefont {Silberhorn}},\ }\bibfield  {title}
	{\bibinfo {title} {{A 2D Quantum Walk Simulation of Two-Particle Dynamics}},\
	}\href {https://doi.org/10.1126/science.1218448} {\bibfield  {journal}
		{\bibinfo  {journal} {Science}\ }\textbf {\bibinfo {volume} {336}},\ \bibinfo
		{pages} {55} (\bibinfo {year} {2012})}\BibitemShut {NoStop}%
	\bibitem [{\citenamefont {Travaglione}\ and\ \citenamefont
		{Milburn}(2002)}]{Milburn2002}%
	\BibitemOpen
	\bibfield  {author} {\bibinfo {author} {\bibfnamefont {B.~C.}\ \bibnamefont
			{Travaglione}}\ and\ \bibinfo {author} {\bibfnamefont {G.~J.}\ \bibnamefont
			{Milburn}},\ }\bibfield  {title} {\bibinfo {title} {Implementing the quantum
			random walk},\ }\href {https://doi.org/10.1103/PhysRevA.65.032310} {\bibfield
		{journal} {\bibinfo  {journal} {Phys. Rev. A}\ }\textbf {\bibinfo {volume}
			{65}},\ \bibinfo {pages} {032310} (\bibinfo {year} {2002})}\BibitemShut
	{NoStop}%
	\bibitem [{\citenamefont {Schmitz}\ \emph {et~al.}(2009)\citenamefont
		{Schmitz}, \citenamefont {Matjeschk}, \citenamefont {Schneider},
		\citenamefont {Glueckert}, \citenamefont {Enderlein}, \citenamefont {Huber},\
		and\ \citenamefont {Schaetz}}]{Schmitz2009}%
	\BibitemOpen
	\bibfield  {author} {\bibinfo {author} {\bibfnamefont {H.}~\bibnamefont
			{Schmitz}}, \bibinfo {author} {\bibfnamefont {R.}~\bibnamefont {Matjeschk}},
		\bibinfo {author} {\bibfnamefont {C.}~\bibnamefont {Schneider}}, \bibinfo
		{author} {\bibfnamefont {J.}~\bibnamefont {Glueckert}}, \bibinfo {author}
		{\bibfnamefont {M.}~\bibnamefont {Enderlein}}, \bibinfo {author}
		{\bibfnamefont {T.}~\bibnamefont {Huber}},\ and\ \bibinfo {author}
		{\bibfnamefont {T.}~\bibnamefont {Schaetz}},\ }\bibfield  {title} {\bibinfo
		{title} {Quantum walk of a trapped ion in phase space},\ }\href
	{https://doi.org/10.1103/PhysRevLett.103.090504} {\bibfield  {journal}
		{\bibinfo  {journal} {Phys. Rev. Lett.}\ }\textbf {\bibinfo {volume} {103}},\
		\bibinfo {pages} {090504} (\bibinfo {year} {2009})}\BibitemShut {NoStop}%
	\bibitem [{\citenamefont {Karski}\ \emph {et~al.}(2009)\citenamefont {Karski},
		\citenamefont {F{\"o}rster}, \citenamefont {Choi}, \citenamefont {Steffen},
		\citenamefont {Alt}, \citenamefont {Meschede},\ and\ \citenamefont
		{Widera}}]{Karski2009}%
	\BibitemOpen
	\bibfield  {author} {\bibinfo {author} {\bibfnamefont {M.}~\bibnamefont
			{Karski}}, \bibinfo {author} {\bibfnamefont {L.}~\bibnamefont {F{\"o}rster}},
		\bibinfo {author} {\bibfnamefont {J.-M.}\ \bibnamefont {Choi}}, \bibinfo
		{author} {\bibfnamefont {A.}~\bibnamefont {Steffen}}, \bibinfo {author}
		{\bibfnamefont {W.}~\bibnamefont {Alt}}, \bibinfo {author} {\bibfnamefont
			{D.}~\bibnamefont {Meschede}},\ and\ \bibinfo {author} {\bibfnamefont
			{A.}~\bibnamefont {Widera}},\ }\bibfield  {title} {\bibinfo {title} {Quantum
			walk in position space with single optically trapped atoms},\ }\href
	{https://doi.org/10.1126/science.1174436} {\bibfield  {journal} {\bibinfo
			{journal} {Science}\ }\textbf {\bibinfo {volume} {325}},\ \bibinfo {pages}
		{174} (\bibinfo {year} {2009})}\BibitemShut {NoStop}%
	\bibitem [{\citenamefont {Yan}\ \emph {et~al.}(2019)\citenamefont {Yan},
		\citenamefont {Zhang}, \citenamefont {Gong}, \citenamefont {Wu},
		\citenamefont {Zheng}, \citenamefont {Li}, \citenamefont {Wang},
		\citenamefont {Liang}, \citenamefont {Lin}, \citenamefont {Xu}, \citenamefont
		{Guo}, \citenamefont {Sun}, \citenamefont {Peng}, \citenamefont {Xia},
		\citenamefont {Deng}, \citenamefont {Rong}, \citenamefont {You},
		\citenamefont {Nori}, \citenamefont {Fan}, \citenamefont {Zhu},\ and\
		\citenamefont {Pan}}]{Pan2019}%
	\BibitemOpen
	\bibfield  {author} {\bibinfo {author} {\bibfnamefont {Z.}~\bibnamefont
			{Yan}}, \bibinfo {author} {\bibfnamefont {Y.-R.}\ \bibnamefont {Zhang}},
		\bibinfo {author} {\bibfnamefont {M.}~\bibnamefont {Gong}}, \bibinfo {author}
		{\bibfnamefont {Y.}~\bibnamefont {Wu}}, \bibinfo {author} {\bibfnamefont
			{Y.}~\bibnamefont {Zheng}}, \bibinfo {author} {\bibfnamefont
			{S.}~\bibnamefont {Li}}, \bibinfo {author} {\bibfnamefont {C.}~\bibnamefont
			{Wang}}, \bibinfo {author} {\bibfnamefont {F.}~\bibnamefont {Liang}},
		\bibinfo {author} {\bibfnamefont {J.}~\bibnamefont {Lin}}, \bibinfo {author}
		{\bibfnamefont {Y.}~\bibnamefont {Xu}}, \bibinfo {author} {\bibfnamefont
			{C.}~\bibnamefont {Guo}}, \bibinfo {author} {\bibfnamefont {L.}~\bibnamefont
			{Sun}}, \bibinfo {author} {\bibfnamefont {C.-Z.}\ \bibnamefont {Peng}},
		\bibinfo {author} {\bibfnamefont {K.}~\bibnamefont {Xia}}, \bibinfo {author}
		{\bibfnamefont {H.}~\bibnamefont {Deng}}, \bibinfo {author} {\bibfnamefont
			{H.}~\bibnamefont {Rong}}, \bibinfo {author} {\bibfnamefont {J.~Q.}\
			\bibnamefont {You}}, \bibinfo {author} {\bibfnamefont {F.}~\bibnamefont
			{Nori}}, \bibinfo {author} {\bibfnamefont {H.}~\bibnamefont {Fan}}, \bibinfo
		{author} {\bibfnamefont {X.}~\bibnamefont {Zhu}},\ and\ \bibinfo {author}
		{\bibfnamefont {J.-W.}\ \bibnamefont {Pan}},\ }\bibfield  {title} {\bibinfo
		{title} {Strongly correlated quantum walks with a 12-qubit superconducting
			processor},\ }\href {https://doi.org/10.1126/science.aaw1611} {\bibfield
		{journal} {\bibinfo  {journal} {Science}\ }\textbf {\bibinfo {volume}
			{364}},\ \bibinfo {pages} {753} (\bibinfo {year} {2019})}\BibitemShut
	{NoStop}%
	\bibitem [{\citenamefont {Gong}\ \emph {et~al.}(2021)\citenamefont {Gong},
		\citenamefont {Wang}, \citenamefont {Zha}, \citenamefont {Chen},
		\citenamefont {Huang}, \citenamefont {Wu}, \citenamefont {Zhu}, \citenamefont
		{Zhao}, \citenamefont {Li}, \citenamefont {Guo}, \citenamefont {Qian},
		\citenamefont {Ye}, \citenamefont {Chen}, \citenamefont {Ying}, \citenamefont
		{Yu}, \citenamefont {Fan}, \citenamefont {Wu}, \citenamefont {Su},
		\citenamefont {Deng}, \citenamefont {Rong}, \citenamefont {Zhang},
		\citenamefont {Cao}, \citenamefont {Lin}, \citenamefont {Xu}, \citenamefont
		{Sun}, \citenamefont {Guo}, \citenamefont {Li}, \citenamefont {Liang},
		\citenamefont {Bastidas}, \citenamefont {Nemoto}, \citenamefont {Munro},
		\citenamefont {Huo}, \citenamefont {Lu}, \citenamefont {Peng}, \citenamefont
		{Zhu},\ and\ \citenamefont {Pan}}]{Pan2021}%
	\BibitemOpen
	\bibfield  {author} {\bibinfo {author} {\bibfnamefont {M.}~\bibnamefont
			{Gong}}, \bibinfo {author} {\bibfnamefont {S.}~\bibnamefont {Wang}}, \bibinfo
		{author} {\bibfnamefont {C.}~\bibnamefont {Zha}}, \bibinfo {author}
		{\bibfnamefont {M.-C.}\ \bibnamefont {Chen}}, \bibinfo {author}
		{\bibfnamefont {H.-L.}\ \bibnamefont {Huang}}, \bibinfo {author}
		{\bibfnamefont {Y.}~\bibnamefont {Wu}}, \bibinfo {author} {\bibfnamefont
			{Q.}~\bibnamefont {Zhu}}, \bibinfo {author} {\bibfnamefont {Y.}~\bibnamefont
			{Zhao}}, \bibinfo {author} {\bibfnamefont {S.}~\bibnamefont {Li}}, \bibinfo
		{author} {\bibfnamefont {S.}~\bibnamefont {Guo}}, \bibinfo {author}
		{\bibfnamefont {H.}~\bibnamefont {Qian}}, \bibinfo {author} {\bibfnamefont
			{Y.}~\bibnamefont {Ye}}, \bibinfo {author} {\bibfnamefont {F.}~\bibnamefont
			{Chen}}, \bibinfo {author} {\bibfnamefont {C.}~\bibnamefont {Ying}}, \bibinfo
		{author} {\bibfnamefont {J.}~\bibnamefont {Yu}}, \bibinfo {author}
		{\bibfnamefont {D.}~\bibnamefont {Fan}}, \bibinfo {author} {\bibfnamefont
			{D.}~\bibnamefont {Wu}}, \bibinfo {author} {\bibfnamefont {H.}~\bibnamefont
			{Su}}, \bibinfo {author} {\bibfnamefont {H.}~\bibnamefont {Deng}}, \bibinfo
		{author} {\bibfnamefont {H.}~\bibnamefont {Rong}}, \bibinfo {author}
		{\bibfnamefont {K.}~\bibnamefont {Zhang}}, \bibinfo {author} {\bibfnamefont
			{S.}~\bibnamefont {Cao}}, \bibinfo {author} {\bibfnamefont {J.}~\bibnamefont
			{Lin}}, \bibinfo {author} {\bibfnamefont {Y.}~\bibnamefont {Xu}}, \bibinfo
		{author} {\bibfnamefont {L.}~\bibnamefont {Sun}}, \bibinfo {author}
		{\bibfnamefont {C.}~\bibnamefont {Guo}}, \bibinfo {author} {\bibfnamefont
			{N.}~\bibnamefont {Li}}, \bibinfo {author} {\bibfnamefont {F.}~\bibnamefont
			{Liang}}, \bibinfo {author} {\bibfnamefont {V.~M.}\ \bibnamefont {Bastidas}},
		\bibinfo {author} {\bibfnamefont {K.}~\bibnamefont {Nemoto}}, \bibinfo
		{author} {\bibfnamefont {W.~J.}\ \bibnamefont {Munro}}, \bibinfo {author}
		{\bibfnamefont {Y.-H.}\ \bibnamefont {Huo}}, \bibinfo {author} {\bibfnamefont
			{C.-Y.}\ \bibnamefont {Lu}}, \bibinfo {author} {\bibfnamefont {C.-Z.}\
			\bibnamefont {Peng}}, \bibinfo {author} {\bibfnamefont {X.}~\bibnamefont
			{Zhu}},\ and\ \bibinfo {author} {\bibfnamefont {J.-W.}\ \bibnamefont {Pan}},\
	}\bibfield  {title} {\bibinfo {title} {Quantum walks on a programmable
			two-dimensional 62-qubit superconducting processor},\ }\href
	{https://doi.org/10.1126/science.abg7812} {\bibfield  {journal} {\bibinfo
			{journal} {Science}\ }\textbf {\bibinfo {volume} {372}},\ \bibinfo {pages}
		{948} (\bibinfo {year} {2021})}\BibitemShut {NoStop}%
	\bibitem [{\citenamefont {Kang}\ \emph {et~al.}(2025)\citenamefont {Kang},
		\citenamefont {Su}, \citenamefont {Wang}, \citenamefont {Shen},\ and\
		\citenamefont {Yang}}]{Yang2025}%
	\BibitemOpen
	\bibfield  {author} {\bibinfo {author} {\bibfnamefont {Y.-H.}\ \bibnamefont
			{Kang}}, \bibinfo {author} {\bibfnamefont {Q.-P.}\ \bibnamefont {Su}},
		\bibinfo {author} {\bibfnamefont {Y.}~\bibnamefont {Wang}}, \bibinfo {author}
		{\bibfnamefont {L.}~\bibnamefont {Shen}},\ and\ \bibinfo {author}
		{\bibfnamefont {C.-P.}\ \bibnamefont {Yang}},\ }\bibfield  {title} {\bibinfo
		{title} {Investigation of the topological quantum phase transition in a
			superconducting system via quantum walks},\ }\href
	{https://doi.org/10.1103/7dy8-1g4t} {\bibfield  {journal} {\bibinfo
			{journal} {Phys. Rev. A}\ }\textbf {\bibinfo {volume} {111}},\ \bibinfo
		{pages} {062209} (\bibinfo {year} {2025})}\BibitemShut {NoStop}%
	\bibitem [{\citenamefont {Manouchehri}\ and\ \citenamefont
		{Wang}(2014)}]{Manouchehri2014}%
	\BibitemOpen
	\bibfield  {author} {\bibinfo {author} {\bibfnamefont {K.}~\bibnamefont
			{Manouchehri}}\ and\ \bibinfo {author} {\bibfnamefont {J.}~\bibnamefont
			{Wang}},\ }\bibinfo {title} {Physical implementation},\ in\ \href
	{https://doi.org/10.1007/978-3-642-36014-5_3} {\emph {\bibinfo {booktitle}
			{Physical Implementation of Quantum Walks}}}\ (\bibinfo  {publisher}
	{Springer Berlin Heidelberg},\ \bibinfo {address} {Berlin, Heidelberg},\
	\bibinfo {year} {2014})\ pp.\ \bibinfo {pages} {39--150}\BibitemShut
	{NoStop}%
	\bibitem [{\citenamefont {Vieira}\ \emph {et~al.}(2013)\citenamefont {Vieira},
		\citenamefont {Amorim},\ and\ \citenamefont {Rigolin}}]{Vieira2013}%
	\BibitemOpen
	\bibfield  {author} {\bibinfo {author} {\bibfnamefont {R.}~\bibnamefont
			{Vieira}}, \bibinfo {author} {\bibfnamefont {E.~P.~M.}\ \bibnamefont
			{Amorim}},\ and\ \bibinfo {author} {\bibfnamefont {G.}~\bibnamefont
			{Rigolin}},\ }\bibfield  {title} {\bibinfo {title} {Dynamically disordered
			quantum walk as a maximal entanglement generator},\ }\href
	{https://doi.org/10.1103/PhysRevLett.111.180503} {\bibfield  {journal}
		{\bibinfo  {journal} {Phys. Rev. Lett.}\ }\textbf {\bibinfo {volume} {111}},\
		\bibinfo {pages} {180503} (\bibinfo {year} {2013})}\BibitemShut {NoStop}%
	\bibitem [{\citenamefont {Singh}\ \emph {et~al.}(2025)\citenamefont {Singh},
		\citenamefont {Mittal},\ and\ \citenamefont {Bose}}]{Mittal2025}%
	\BibitemOpen
	\bibfield  {author} {\bibinfo {author} {\bibfnamefont {J.}~\bibnamefont
			{Singh}}, \bibinfo {author} {\bibfnamefont {V.}~\bibnamefont {Mittal}},\ and\
		\bibinfo {author} {\bibfnamefont {S.}~\bibnamefont {Bose}},\ }\bibfield
	{title} {\bibinfo {title} {Deterministic generation of hybrid entangled
			states using quantum walks},\ }\href
	{https://doi.org/10.1007/s11128-025-04886-4} {\bibfield  {journal} {\bibinfo
			{journal} {Quantum Information Processing}\ }\textbf {\bibinfo {volume}
			{24}},\ \bibinfo {pages} {269} (\bibinfo {year} {2025})}\BibitemShut
	{NoStop}%
	\bibitem [{\citenamefont {Robens}\ \emph {et~al.}(2015)\citenamefont {Robens},
		\citenamefont {Alt}, \citenamefont {Meschede}, \citenamefont {Emary},\ and\
		\citenamefont {Alberti}}]{Alberti2015}%
	\BibitemOpen
	\bibfield  {author} {\bibinfo {author} {\bibfnamefont {C.}~\bibnamefont
			{Robens}}, \bibinfo {author} {\bibfnamefont {W.}~\bibnamefont {Alt}},
		\bibinfo {author} {\bibfnamefont {D.}~\bibnamefont {Meschede}}, \bibinfo
		{author} {\bibfnamefont {C.}~\bibnamefont {Emary}},\ and\ \bibinfo {author}
		{\bibfnamefont {A.}~\bibnamefont {Alberti}},\ }\bibfield  {title} {\bibinfo
		{title} {Ideal negative measurements in quantum walks disprove theories based
			on classical trajectories},\ }\href
	{https://doi.org/10.1103/PhysRevX.5.011003} {\bibfield  {journal} {\bibinfo
			{journal} {Phys. Rev. X}\ }\textbf {\bibinfo {volume} {5}},\ \bibinfo {pages}
		{011003} (\bibinfo {year} {2015})}\BibitemShut {NoStop}%
	\bibitem [{\citenamefont {Orthey}\ and\ \citenamefont
		{Angelo}(2019)}]{Angelo2019}%
	\BibitemOpen
	\bibfield  {author} {\bibinfo {author} {\bibfnamefont {A.~C.}\ \bibnamefont
			{Orthey}}\ and\ \bibinfo {author} {\bibfnamefont {R.~M.}\ \bibnamefont
			{Angelo}},\ }\bibfield  {title} {\bibinfo {title} {Nonlocality, quantum
			correlations, and violations of classical realism in the dynamics of two
			noninteracting quantum walkers},\ }\href
	{https://doi.org/10.1103/PhysRevA.100.042110} {\bibfield  {journal} {\bibinfo
			{journal} {Phys. Rev. A}\ }\textbf {\bibinfo {volume} {100}},\ \bibinfo
		{pages} {042110} (\bibinfo {year} {2019})}\BibitemShut {NoStop}%
	\bibitem [{\citenamefont {Kadian}\ \emph {et~al.}(2021)\citenamefont {Kadian},
		\citenamefont {Garhwal},\ and\ \citenamefont {Kumar}}]{Kadian2021}%
	\BibitemOpen
	\bibfield  {author} {\bibinfo {author} {\bibfnamefont {K.}~\bibnamefont
			{Kadian}}, \bibinfo {author} {\bibfnamefont {S.}~\bibnamefont {Garhwal}},\
		and\ \bibinfo {author} {\bibfnamefont {A.}~\bibnamefont {Kumar}},\ }\bibfield
	{title} {\bibinfo {title} {Quantum walk and its application domains: A
			systematic review},\ }\href
	{https://doi.org/https://doi.org/10.1016/j.cosrev.2021.100419} {\bibfield
		{journal} {\bibinfo  {journal} {Computer Science Review}\ }\textbf {\bibinfo
			{volume} {41}},\ \bibinfo {pages} {100419} (\bibinfo {year}
		{2021})}\BibitemShut {NoStop}%
	\bibitem [{\citenamefont {{Pa{\c{s}}cu Moca}}\ \emph
		{et~al.}(2025)\citenamefont {{Pa{\c{s}}cu Moca}}, \citenamefont {{Sticlet}},
		\citenamefont {{D{\'o}ra}}, \citenamefont {{Valli}}, \citenamefont
		{{Szombathy}},\ and\ \citenamefont {{Zar{\'a}nd}}}]{QWMagic2025}%
	\BibitemOpen
	\bibfield  {author} {\bibinfo {author} {\bibfnamefont {C.}~\bibnamefont
			{{Pa{\c{s}}cu Moca}}}, \bibinfo {author} {\bibfnamefont {D.}~\bibnamefont
			{{Sticlet}}}, \bibinfo {author} {\bibfnamefont {B.}~\bibnamefont
			{{D{\'o}ra}}}, \bibinfo {author} {\bibfnamefont {A.}~\bibnamefont {{Valli}}},
		\bibinfo {author} {\bibfnamefont {D.}~\bibnamefont {{Szombathy}}},\ and\
		\bibinfo {author} {\bibfnamefont {G.}~\bibnamefont {{Zar{\'a}nd}}},\
	}\bibfield  {title} {\bibinfo {title} {{Non-stabilizerness generation in a
				multi-particle quantum walk}},\ }\href
	{https://doi.org/10.48550/arXiv.2504.19750} {\bibfield  {journal} {\bibinfo
			{journal} {arXiv e-prints}\ ,\ \bibinfo {eid} {arXiv:2504.19750}} (\bibinfo
		{year} {2025})}\BibitemShut {NoStop}%
	\bibitem [{\citenamefont {Werner}(1989)}]{Werner1989}%
	\BibitemOpen
	\bibfield  {author} {\bibinfo {author} {\bibfnamefont {R.~F.}\ \bibnamefont
			{Werner}},\ }\bibfield  {title} {\bibinfo {title} {{Quantum states with
				Einstein-Podolsky-Rosen correlations admitting a hidden-variable model}},\
	}\href {https://doi.org/10.1103/PhysRevA.40.4277} {\bibfield  {journal}
		{\bibinfo  {journal} {Phys. Rev. A}\ }\textbf {\bibinfo {volume} {40}},\
		\bibinfo {pages} {4277} (\bibinfo {year} {1989})}\BibitemShut {NoStop}%
	\bibitem [{\citenamefont {Wang}\ \emph {et~al.}(2020)\citenamefont {Wang},
		\citenamefont {Xu}, \citenamefont {Pan}, \citenamefont {Tao}, \citenamefont
		{Chen}, \citenamefont {Zhan}, \citenamefont {Sun}, \citenamefont {Xu},
		\citenamefont {Chen}, \citenamefont {Han}, \citenamefont {Li},\ and\
		\citenamefont {Guo}}]{Wang2020}%
	\BibitemOpen
	\bibfield  {author} {\bibinfo {author} {\bibfnamefont {Q.-Q.}\ \bibnamefont
			{Wang}}, \bibinfo {author} {\bibfnamefont {X.-Y.}\ \bibnamefont {Xu}},
		\bibinfo {author} {\bibfnamefont {W.-W.}\ \bibnamefont {Pan}}, \bibinfo
		{author} {\bibfnamefont {S.-J.}\ \bibnamefont {Tao}}, \bibinfo {author}
		{\bibfnamefont {Z.}~\bibnamefont {Chen}}, \bibinfo {author} {\bibfnamefont
			{Y.-T.}\ \bibnamefont {Zhan}}, \bibinfo {author} {\bibfnamefont
			{K.}~\bibnamefont {Sun}}, \bibinfo {author} {\bibfnamefont {J.-S.}\
			\bibnamefont {Xu}}, \bibinfo {author} {\bibfnamefont {G.}~\bibnamefont
			{Chen}}, \bibinfo {author} {\bibfnamefont {Y.-J.}\ \bibnamefont {Han}},
		\bibinfo {author} {\bibfnamefont {C.-F.}\ \bibnamefont {Li}},\ and\ \bibinfo
		{author} {\bibfnamefont {G.-C.}\ \bibnamefont {Guo}},\ }\bibfield  {title}
	{\bibinfo {title} {Robustness of entanglement as an indicator of topological
			phases in quantum walks},\ }\href {https://doi.org/10.1364/OPTICA.375388}
	{\bibfield  {journal} {\bibinfo  {journal} {Optica}\ }\textbf {\bibinfo
			{volume} {7}},\ \bibinfo {pages} {53} (\bibinfo {year} {2020})}\BibitemShut
	{NoStop}%
	\bibitem [{\citenamefont {{Nayak}}\ and\ \citenamefont
		{{Vishwanath}}(2000)}]{Nayak2000}%
	\BibitemOpen
	\bibfield  {author} {\bibinfo {author} {\bibfnamefont {A.}~\bibnamefont
			{{Nayak}}}\ and\ \bibinfo {author} {\bibfnamefont {A.}~\bibnamefont
			{{Vishwanath}}},\ }\bibfield  {title} {\bibinfo {title} {{Quantum Walk on the
				Line}},\ }\href@noop {} {\bibfield  {journal} {\bibinfo  {journal} {arXiv
				e-prints}\ ,\ \bibinfo {eid} {quant-ph/0010117}} (\bibinfo {year}
		{2000})}\BibitemShut {NoStop}%
	\bibitem [{\citenamefont {Ambainis}(2003)}]{Ambainis2003}%
	\BibitemOpen
	\bibfield  {author} {\bibinfo {author} {\bibfnamefont {A.}~\bibnamefont
			{Ambainis}},\ }\bibfield  {title} {\bibinfo {title} {Quantum walks and their
			algorithmic applications},\ }\href
	{https://doi.org/10.1142/S0219749903000383} {\bibfield  {journal} {\bibinfo
			{journal} {Int. J. Quantum Inf.}\ }\textbf {\bibinfo {volume} {01}},\
		\bibinfo {pages} {507} (\bibinfo {year} {2003})}\BibitemShut {NoStop}%
	\bibitem [{\citenamefont {Childs}\ and\ \citenamefont
		{Goldstone}(2004)}]{Childs2004}%
	\BibitemOpen
	\bibfield  {author} {\bibinfo {author} {\bibfnamefont {A.~M.}\ \bibnamefont
			{Childs}}\ and\ \bibinfo {author} {\bibfnamefont {J.}~\bibnamefont
			{Goldstone}},\ }\bibfield  {title} {\bibinfo {title} {Spatial search by
			quantum walk},\ }\href {https://doi.org/10.1103/PhysRevA.70.022314}
	{\bibfield  {journal} {\bibinfo  {journal} {Phys. Rev. A}\ }\textbf {\bibinfo
			{volume} {70}},\ \bibinfo {pages} {022314} (\bibinfo {year}
		{2004})}\BibitemShut {NoStop}%
	\bibitem [{\citenamefont {Shenvi}\ \emph {et~al.}(2003)\citenamefont {Shenvi},
		\citenamefont {Kempe},\ and\ \citenamefont {Whaley}}]{Shenvi2003}%
	\BibitemOpen
	\bibfield  {author} {\bibinfo {author} {\bibfnamefont {N.}~\bibnamefont
			{Shenvi}}, \bibinfo {author} {\bibfnamefont {J.}~\bibnamefont {Kempe}},\ and\
		\bibinfo {author} {\bibfnamefont {K.~B.}\ \bibnamefont {Whaley}},\ }\bibfield
	{title} {\bibinfo {title} {Quantum random-walk search algorithm},\ }\href
	{https://doi.org/10.1103/PhysRevA.67.052307} {\bibfield  {journal} {\bibinfo
			{journal} {Phys. Rev. A}\ }\textbf {\bibinfo {volume} {67}},\ \bibinfo
		{pages} {052307} (\bibinfo {year} {2003})}\BibitemShut {NoStop}%
	\bibitem [{\citenamefont {Childs}(2009)}]{Childs2009}%
	\BibitemOpen
	\bibfield  {author} {\bibinfo {author} {\bibfnamefont {A.~M.}\ \bibnamefont
			{Childs}},\ }\bibfield  {title} {\bibinfo {title} {Universal computation by
			quantum walk},\ }\href {https://doi.org/10.1103/PhysRevLett.102.180501}
	{\bibfield  {journal} {\bibinfo  {journal} {Phys. Rev. Lett.}\ }\textbf
		{\bibinfo {volume} {102}},\ \bibinfo {pages} {180501} (\bibinfo {year}
		{2009})}\BibitemShut {NoStop}%
	\bibitem [{\citenamefont {Childs}\ \emph {et~al.}(2013)\citenamefont {Childs},
		\citenamefont {Gosset},\ and\ \citenamefont {Webb}}]{Childs2013}%
	\BibitemOpen
	\bibfield  {author} {\bibinfo {author} {\bibfnamefont {A.~M.}\ \bibnamefont
			{Childs}}, \bibinfo {author} {\bibfnamefont {D.}~\bibnamefont {Gosset}},\
		and\ \bibinfo {author} {\bibfnamefont {Z.}~\bibnamefont {Webb}},\ }\bibfield
	{title} {\bibinfo {title} {Universal computation by multiparticle quantum
			walk},\ }\href {https://doi.org/10.1126/science.1229957} {\bibfield
		{journal} {\bibinfo  {journal} {Science}\ }\textbf {\bibinfo {volume}
			{339}},\ \bibinfo {pages} {791} (\bibinfo {year} {2013})}\BibitemShut
	{NoStop}%
	\bibitem [{\citenamefont {Lovett}\ \emph {et~al.}(2010)\citenamefont {Lovett},
		\citenamefont {Cooper}, \citenamefont {Everitt}, \citenamefont {Trevers},\
		and\ \citenamefont {Kendon}}]{Lovett2010}%
	\BibitemOpen
	\bibfield  {author} {\bibinfo {author} {\bibfnamefont {N.~B.}\ \bibnamefont
			{Lovett}}, \bibinfo {author} {\bibfnamefont {S.}~\bibnamefont {Cooper}},
		\bibinfo {author} {\bibfnamefont {M.}~\bibnamefont {Everitt}}, \bibinfo
		{author} {\bibfnamefont {M.}~\bibnamefont {Trevers}},\ and\ \bibinfo {author}
		{\bibfnamefont {V.}~\bibnamefont {Kendon}},\ }\bibfield  {title} {\bibinfo
		{title} {Universal quantum computation using the discrete-time quantum
			walk},\ }\href {https://doi.org/10.1103/PhysRevA.81.042330} {\bibfield
		{journal} {\bibinfo  {journal} {Phys. Rev. A}\ }\textbf {\bibinfo {volume}
			{81}},\ \bibinfo {pages} {042330} (\bibinfo {year} {2010})}\BibitemShut
	{NoStop}%
	\bibitem [{\citenamefont {Venegas-Andraca}(2012)}]{VenegasAndraca2012}%
	\BibitemOpen
	\bibfield  {author} {\bibinfo {author} {\bibfnamefont {S.~E.}\ \bibnamefont
			{Venegas-Andraca}},\ }\bibfield  {title} {\bibinfo {title} {Quantum walks: a
			comprehensive review},\ }\href {https://doi.org/10.1007/s11128-012-0432-5}
	{\bibfield  {journal} {\bibinfo  {journal} {Quantum Inf. Process.}\ }\textbf
		{\bibinfo {volume} {11}},\ \bibinfo {pages} {1015} (\bibinfo {year}
		{2012})}\BibitemShut {NoStop}%
	\bibitem [{\citenamefont {Leone}\ \emph {et~al.}(2022)\citenamefont {Leone},
		\citenamefont {Oliviero},\ and\ \citenamefont {Hamma}}]{Leone2022}%
	\BibitemOpen
	\bibfield  {author} {\bibinfo {author} {\bibfnamefont {L.}~\bibnamefont
			{Leone}}, \bibinfo {author} {\bibfnamefont {S.~F.~E.}\ \bibnamefont
			{Oliviero}},\ and\ \bibinfo {author} {\bibfnamefont {A.}~\bibnamefont
			{Hamma}},\ }\bibfield  {title} {\bibinfo {title} {Stabilizer r\'enyi
			entropy},\ }\href {https://doi.org/10.1103/PhysRevLett.128.050402} {\bibfield
		{journal} {\bibinfo  {journal} {Phys. Rev. Lett.}\ }\textbf {\bibinfo
			{volume} {128}},\ \bibinfo {pages} {050402} (\bibinfo {year}
		{2022})}\BibitemShut {NoStop}%
	\bibitem [{\citenamefont {Haug}\ and\ \citenamefont {Piroli}(2023)}]{Haug2023}%
	\BibitemOpen
	\bibfield  {author} {\bibinfo {author} {\bibfnamefont {T.}~\bibnamefont
			{Haug}}\ and\ \bibinfo {author} {\bibfnamefont {L.}~\bibnamefont {Piroli}},\
	}\bibfield  {title} {\bibinfo {title} {Stabilizer entropies and
			nonstabilizerness monotones},\ }\href
	{https://doi.org/10.22331/q-2023-08-28-1092} {\bibfield  {journal} {\bibinfo
			{journal} {{Quantum}}\ }\textbf {\bibinfo {volume} {7}},\ \bibinfo {pages}
		{1092} (\bibinfo {year} {2023})}\BibitemShut {NoStop}%
	\bibitem [{\citenamefont {Leone}\ and\ \citenamefont
		{Bittel}(2024)}]{Leone2024}%
	\BibitemOpen
	\bibfield  {author} {\bibinfo {author} {\bibfnamefont {L.}~\bibnamefont
			{Leone}}\ and\ \bibinfo {author} {\bibfnamefont {L.}~\bibnamefont {Bittel}},\
	}\bibfield  {title} {\bibinfo {title} {Stabilizer entropies are monotones for
			magic-state resource theory},\ }\href
	{https://doi.org/10.1103/PhysRevA.110.L040403} {\bibfield  {journal}
		{\bibinfo  {journal} {Phys. Rev. A}\ }\textbf {\bibinfo {volume} {110}},\
		\bibinfo {pages} {L040403} (\bibinfo {year} {2024})}\BibitemShut {NoStop}%
	\bibitem [{\citenamefont {Oliviero}\ \emph {et~al.}(2022)\citenamefont
		{Oliviero}, \citenamefont {Leone}, \citenamefont {Hamma},\ and\ \citenamefont
		{Lloyd}}]{Oliviero2022}%
	\BibitemOpen
	\bibfield  {author} {\bibinfo {author} {\bibfnamefont {S.~F.~E.}\
			\bibnamefont {Oliviero}}, \bibinfo {author} {\bibfnamefont {L.}~\bibnamefont
			{Leone}}, \bibinfo {author} {\bibfnamefont {A.}~\bibnamefont {Hamma}},\ and\
		\bibinfo {author} {\bibfnamefont {S.}~\bibnamefont {Lloyd}},\ }\bibfield
	{title} {\bibinfo {title} {Measuring magic on a quantum processor},\ }\href
	{https://doi.org/10.1038/s41534-022-00666-5} {\bibfield  {journal} {\bibinfo
			{journal} {npj Quantum Information}\ }\textbf {\bibinfo {volume} {8}},\
		\bibinfo {pages} {148} (\bibinfo {year} {2022})}\BibitemShut {NoStop}%
	\bibitem [{\citenamefont {Haug}\ \emph {et~al.}(2024)\citenamefont {Haug},
		\citenamefont {Lee},\ and\ \citenamefont {Kim}}]{Kim2024}%
	\BibitemOpen
	\bibfield  {author} {\bibinfo {author} {\bibfnamefont {T.}~\bibnamefont
			{Haug}}, \bibinfo {author} {\bibfnamefont {S.}~\bibnamefont {Lee}},\ and\
		\bibinfo {author} {\bibfnamefont {M.~S.}\ \bibnamefont {Kim}},\ }\bibfield
	{title} {\bibinfo {title} {Efficient quantum algorithms for stabilizer
			entropies},\ }\href {https://doi.org/10.1103/PhysRevLett.132.240602}
	{\bibfield  {journal} {\bibinfo  {journal} {Phys. Rev. Lett.}\ }\textbf
		{\bibinfo {volume} {132}},\ \bibinfo {pages} {240602} (\bibinfo {year}
		{2024})}\BibitemShut {NoStop}%
	\bibitem [{\citenamefont {Wang}\ and\ \citenamefont {Li}(2023)}]{Wang2023}%
	\BibitemOpen
	\bibfield  {author} {\bibinfo {author} {\bibfnamefont {Y.}~\bibnamefont
			{Wang}}\ and\ \bibinfo {author} {\bibfnamefont {Y.}~\bibnamefont {Li}},\
	}\bibfield  {title} {\bibinfo {title} {{Stabilizer R{\'e}nyi entropy on
				qudits}},\ }\href {https://doi.org/10.1007/s11128-023-04186-9} {\bibfield
		{journal} {\bibinfo  {journal} {Quantum Information Processing}\ }\textbf
		{\bibinfo {volume} {22}},\ \bibinfo {pages} {444} (\bibinfo {year}
		{2023})}\BibitemShut {NoStop}%
	\bibitem [{\citenamefont {{Cuffaro}}\ and\ \citenamefont
		{{Fuchs}}(2024)}]{Fuchs2024}%
	\BibitemOpen
	\bibfield  {author} {\bibinfo {author} {\bibfnamefont {G.}~\bibnamefont
			{{Cuffaro}}}\ and\ \bibinfo {author} {\bibfnamefont {C.~A.}\ \bibnamefont
			{{Fuchs}}},\ }\bibfield  {title} {\bibinfo {title} {{Quantum States with
				Maximal Magic}},\ }\href {https://doi.org/10.48550/arXiv.2412.21083}
	{\bibfield  {journal} {\bibinfo  {journal} {arXiv e-prints}\ ,\ \bibinfo
			{eid} {arXiv:2412.21083}} (\bibinfo {year} {2024})}\BibitemShut {NoStop}%
	\bibitem [{\citenamefont {{Liu}}\ \emph {et~al.}(2025)\citenamefont {{Liu}},
		\citenamefont {{Low}},\ and\ \citenamefont {{Yin}}}]{Zhewei2025}%
	\BibitemOpen
	\bibfield  {author} {\bibinfo {author} {\bibfnamefont {Q.}~\bibnamefont
			{{Liu}}}, \bibinfo {author} {\bibfnamefont {I.}~\bibnamefont {{Low}}},\ and\
		\bibinfo {author} {\bibfnamefont {Z.}~\bibnamefont {{Yin}}},\ }\bibfield
	{title} {\bibinfo {title} {{{Maximal Magic for Two-qubit States}}},\ }\href
	{https://doi.org/10.48550/arXiv.2502.17550} {\bibfield  {journal} {\bibinfo
			{journal} {arXiv e-prints}\ ,\ \bibinfo {eid} {arXiv:2502.17550}} (\bibinfo
		{year} {2025})}\BibitemShut {NoStop}%
	\bibitem [{\citenamefont {Portugal}(2013)}]{Renato2013}%
	\BibitemOpen
	\bibfield  {author} {\bibinfo {author} {\bibfnamefont {R.}~\bibnamefont
			{Portugal}},\ }\href@noop {} {\emph {\bibinfo {title} {Quantum Walks and
				Search Algorithms}}}\ (\bibinfo  {publisher} {Springer Publishing Company,
		Incorporated},\ \bibinfo {year} {2013})\BibitemShut {NoStop}%
	\bibitem [{\citenamefont {Carneiro}\ \emph {et~al.}(2005)\citenamefont
		{Carneiro}, \citenamefont {Loo}, \citenamefont {Xu}, \citenamefont {Girerd},
		\citenamefont {Kendon},\ and\ \citenamefont {Knight}}]{Carneiro2005}%
	\BibitemOpen
	\bibfield  {author} {\bibinfo {author} {\bibfnamefont {I.}~\bibnamefont
			{Carneiro}}, \bibinfo {author} {\bibfnamefont {M.}~\bibnamefont {Loo}},
		\bibinfo {author} {\bibfnamefont {X.}~\bibnamefont {Xu}}, \bibinfo {author}
		{\bibfnamefont {M.}~\bibnamefont {Girerd}}, \bibinfo {author} {\bibfnamefont
			{V.}~\bibnamefont {Kendon}},\ and\ \bibinfo {author} {\bibfnamefont {P.~L.}\
			\bibnamefont {Knight}},\ }\bibfield  {title} {\bibinfo {title} {Entanglement
			in coined quantum walks on regular graphs},\ }\href
	{https://doi.org/10.1088/1367-2630/7/1/156} {\bibfield  {journal} {\bibinfo
			{journal} {New Journal of Physics}\ }\textbf {\bibinfo {volume} {7}},\
		\bibinfo {pages} {156} (\bibinfo {year} {2005})}\BibitemShut {NoStop}%
	\bibitem [{\citenamefont {Abal}\ \emph
		{et~al.}(2006{\natexlab{a}})\citenamefont {Abal}, \citenamefont {Siri},
		\citenamefont {Romanelli},\ and\ \citenamefont {Donangelo}}]{Abal2006}%
	\BibitemOpen
	\bibfield  {author} {\bibinfo {author} {\bibfnamefont {G.}~\bibnamefont
			{Abal}}, \bibinfo {author} {\bibfnamefont {R.}~\bibnamefont {Siri}}, \bibinfo
		{author} {\bibfnamefont {A.}~\bibnamefont {Romanelli}},\ and\ \bibinfo
		{author} {\bibfnamefont {R.}~\bibnamefont {Donangelo}},\ }\bibfield  {title}
	{\bibinfo {title} {Quantum walk on the line: Entanglement and nonlocal
			initial conditions},\ }\href {https://doi.org/10.1103/PhysRevA.73.042302}
	{\bibfield  {journal} {\bibinfo  {journal} {Phys. Rev. A}\ }\textbf {\bibinfo
			{volume} {73}},\ \bibinfo {pages} {042302} (\bibinfo {year}
		{2006}{\natexlab{a}})}\BibitemShut {NoStop}%
	\bibitem [{\citenamefont {Abal}\ \emph
		{et~al.}(2006{\natexlab{b}})\citenamefont {Abal}, \citenamefont {Siri},
		\citenamefont {Romanelli},\ and\ \citenamefont {Donangelo}}]{Abal2006a}%
	\BibitemOpen
	\bibfield  {author} {\bibinfo {author} {\bibfnamefont {G.}~\bibnamefont
			{Abal}}, \bibinfo {author} {\bibfnamefont {R.}~\bibnamefont {Siri}}, \bibinfo
		{author} {\bibfnamefont {A.}~\bibnamefont {Romanelli}},\ and\ \bibinfo
		{author} {\bibfnamefont {R.}~\bibnamefont {Donangelo}},\ }\bibfield  {title}
	{\bibinfo {title} {Erratum: Quantum walk on the line: Entanglement and
			non-local initial conditions {[Phys. Rev. A 73, 042302 (2006)]}},\ }\href
	{https://doi.org/10.1103/PhysRevA.73.069905} {\bibfield  {journal} {\bibinfo
			{journal} {Phys. Rev. A}\ }\textbf {\bibinfo {volume} {73}},\ \bibinfo
		{pages} {069905} (\bibinfo {year} {2006}{\natexlab{b}})}\BibitemShut
	{NoStop}%
	\bibitem [{\citenamefont {Nielsen}\ and\ \citenamefont
		{Chuang}(2010)}]{Nielsen_Chuang_2010}%
	\BibitemOpen
	\bibfield  {author} {\bibinfo {author} {\bibfnamefont {M.~A.}\ \bibnamefont
			{Nielsen}}\ and\ \bibinfo {author} {\bibfnamefont {I.~L.}\ \bibnamefont
			{Chuang}},\ }\href@noop {} {\emph {\bibinfo {title} {Quantum Computation and
				Quantum Information: 10th Anniversary Edition}}}\ (\bibinfo  {publisher}
	{Cambridge University Press},\ \bibinfo {year} {2010})\BibitemShut {NoStop}%
	\bibitem [{\citenamefont {Tang}\ \emph {et~al.}(2018)\citenamefont {Tang},
		\citenamefont {Lin}, \citenamefont {Feng}, \citenamefont {Chen},
		\citenamefont {Gao}, \citenamefont {Sun}, \citenamefont {Wang}, \citenamefont
		{Lai}, \citenamefont {Xu}, \citenamefont {Wang}, \citenamefont {Qiao},
		\citenamefont {Yang},\ and\ \citenamefont {Jin}}]{Tang2018}%
	\BibitemOpen
	\bibfield  {author} {\bibinfo {author} {\bibfnamefont {H.}~\bibnamefont
			{Tang}}, \bibinfo {author} {\bibfnamefont {X.-F.}\ \bibnamefont {Lin}},
		\bibinfo {author} {\bibfnamefont {Z.}~\bibnamefont {Feng}}, \bibinfo {author}
		{\bibfnamefont {J.-Y.}\ \bibnamefont {Chen}}, \bibinfo {author}
		{\bibfnamefont {J.}~\bibnamefont {Gao}}, \bibinfo {author} {\bibfnamefont
			{K.}~\bibnamefont {Sun}}, \bibinfo {author} {\bibfnamefont {C.-Y.}\
			\bibnamefont {Wang}}, \bibinfo {author} {\bibfnamefont {P.-C.}\ \bibnamefont
			{Lai}}, \bibinfo {author} {\bibfnamefont {X.-Y.}\ \bibnamefont {Xu}},
		\bibinfo {author} {\bibfnamefont {Y.}~\bibnamefont {Wang}}, \bibinfo {author}
		{\bibfnamefont {L.-F.}\ \bibnamefont {Qiao}}, \bibinfo {author}
		{\bibfnamefont {A.-L.}\ \bibnamefont {Yang}},\ and\ \bibinfo {author}
		{\bibfnamefont {X.-M.}\ \bibnamefont {Jin}},\ }\bibfield  {title} {\bibinfo
		{title} {Experimental two-dimensional quantum walk on a photonic chip},\
	}\href {https://doi.org/10.1126/sciadv.aat3174} {\bibfield  {journal}
		{\bibinfo  {journal} {Science Advances}\ }\textbf {\bibinfo {volume} {4}},\
		\bibinfo {pages} {eaat3174} (\bibinfo {year} {2018})}\BibitemShut {NoStop}%
	\bibitem [{\citenamefont {Preiss}\ \emph {et~al.}(2015)\citenamefont {Preiss},
		\citenamefont {Ma}, \citenamefont {Tai}, \citenamefont {Lukin}, \citenamefont
		{Rispoli}, \citenamefont {Zupancic}, \citenamefont {Lahini}, \citenamefont
		{Islam},\ and\ \citenamefont {Greiner}}]{Preiss2015}%
	\BibitemOpen
	\bibfield  {author} {\bibinfo {author} {\bibfnamefont {P.~M.}\ \bibnamefont
			{Preiss}}, \bibinfo {author} {\bibfnamefont {R.}~\bibnamefont {Ma}}, \bibinfo
		{author} {\bibfnamefont {M.~E.}\ \bibnamefont {Tai}}, \bibinfo {author}
		{\bibfnamefont {A.}~\bibnamefont {Lukin}}, \bibinfo {author} {\bibfnamefont
			{M.}~\bibnamefont {Rispoli}}, \bibinfo {author} {\bibfnamefont
			{P.}~\bibnamefont {Zupancic}}, \bibinfo {author} {\bibfnamefont
			{Y.}~\bibnamefont {Lahini}}, \bibinfo {author} {\bibfnamefont
			{R.}~\bibnamefont {Islam}},\ and\ \bibinfo {author} {\bibfnamefont
			{M.}~\bibnamefont {Greiner}},\ }\bibfield  {title} {\bibinfo {title}
		{Strongly correlated quantum walks in optical lattices},\ }\href
	{https://doi.org/10.1126/science.1260364} {\bibfield  {journal} {\bibinfo
			{journal} {Science}\ }\textbf {\bibinfo {volume} {347}},\ \bibinfo {pages}
		{1229} (\bibinfo {year} {2015})}\BibitemShut {NoStop}%
	\bibitem [{\citenamefont {Kitagawa}\ \emph {et~al.}(2010)\citenamefont
		{Kitagawa}, \citenamefont {Rudner}, \citenamefont {Berg},\ and\ \citenamefont
		{Demler}}]{Kitagawa2010}%
	\BibitemOpen
	\bibfield  {author} {\bibinfo {author} {\bibfnamefont {T.}~\bibnamefont
			{Kitagawa}}, \bibinfo {author} {\bibfnamefont {M.~S.}\ \bibnamefont
			{Rudner}}, \bibinfo {author} {\bibfnamefont {E.}~\bibnamefont {Berg}},\ and\
		\bibinfo {author} {\bibfnamefont {E.}~\bibnamefont {Demler}},\ }\bibfield
	{title} {\bibinfo {title} {Exploring topological phases with quantum walks},\
	}\href {https://doi.org/10.1103/PhysRevA.82.033429} {\bibfield  {journal}
		{\bibinfo  {journal} {Phys. Rev. A}\ }\textbf {\bibinfo {volume} {82}},\
		\bibinfo {pages} {033429} (\bibinfo {year} {2010})}\BibitemShut {NoStop}%
	\bibitem [{\citenamefont {Asb\'oth}(2012)}]{Asboth2012}%
	\BibitemOpen
	\bibfield  {author} {\bibinfo {author} {\bibfnamefont {J.~K.}\ \bibnamefont
			{Asb\'oth}},\ }\bibfield  {title} {\bibinfo {title} {Symmetries, topological
			phases, and bound states in the one-dimensional quantum walk},\ }\href
	{https://doi.org/10.1103/PhysRevB.86.195414} {\bibfield  {journal} {\bibinfo
			{journal} {Phys. Rev. B}\ }\textbf {\bibinfo {volume} {86}},\ \bibinfo
		{pages} {195414} (\bibinfo {year} {2012})}\BibitemShut {NoStop}%
	\bibitem [{\citenamefont {Mittal}\ \emph {et~al.}(2021)\citenamefont {Mittal},
		\citenamefont {Raj}, \citenamefont {Dey},\ and\ \citenamefont
		{Goyal}}]{Mittal2021}%
	\BibitemOpen
	\bibfield  {author} {\bibinfo {author} {\bibfnamefont {V.}~\bibnamefont
			{Mittal}}, \bibinfo {author} {\bibfnamefont {A.}~\bibnamefont {Raj}},
		\bibinfo {author} {\bibfnamefont {S.}~\bibnamefont {Dey}},\ and\ \bibinfo
		{author} {\bibfnamefont {S.~K.}\ \bibnamefont {Goyal}},\ }\bibfield  {title}
	{\bibinfo {title} {Persistence of topological phases in non-{H}ermitian
			quantum walks},\ }\href {https://doi.org/10.1038/s41598-021-89441-8}
	{\bibfield  {journal} {\bibinfo  {journal} {Sci. Rep.}\ }\textbf {\bibinfo
			{volume} {11}},\ \bibinfo {pages} {10262} (\bibinfo {year}
		{2021})}\BibitemShut {NoStop}%
	\bibitem [{\citenamefont {{Sticlet}}\ \emph {et~al.}(2025)\citenamefont
		{{Sticlet}}, \citenamefont {{D{\'o}ra}}, \citenamefont {{Szombathy}},
		\citenamefont {{Zar{\'a}nd}},\ and\ \citenamefont {{Pa{\c{s}}cu
				Moca}}}]{Sticlet2025}%
	\BibitemOpen
	\bibfield  {author} {\bibinfo {author} {\bibfnamefont {D.}~\bibnamefont
			{{Sticlet}}}, \bibinfo {author} {\bibfnamefont {B.}~\bibnamefont
			{{D{\'o}ra}}}, \bibinfo {author} {\bibfnamefont {D.}~\bibnamefont
			{{Szombathy}}}, \bibinfo {author} {\bibfnamefont {G.}~\bibnamefont
			{{Zar{\'a}nd}}},\ and\ \bibinfo {author} {\bibfnamefont {C.}~\bibnamefont
			{{Pa{\c{s}}cu Moca}}},\ }\bibfield  {title} {\bibinfo {title}
		{{Non-stabilizerness in open XXZ spin chains: Universal scaling and
				dynamics}},\ }\href {https://doi.org/10.48550/arXiv.2504.11139} {\bibfield
		{journal} {\bibinfo  {journal} {arXiv e-prints}\ ,\ \bibinfo {eid}
			{arXiv:2504.11139}} (\bibinfo {year} {2025})}\BibitemShut {NoStop}%
	\bibitem [{\citenamefont {{Tirrito}}\ \emph {et~al.}(2025)\citenamefont
		{{Tirrito}}, \citenamefont {{Sonya Tarabunga}}, \citenamefont {{Singh
				Bhakuni}}, \citenamefont {{Dalmonte}}, \citenamefont {{Sierant}},\ and\
		\citenamefont {{Turkeshi}}}]{Turkeshi2025a}%
	\BibitemOpen
	\bibfield  {author} {\bibinfo {author} {\bibfnamefont {E.}~\bibnamefont
			{{Tirrito}}}, \bibinfo {author} {\bibfnamefont {P.}~\bibnamefont {{Sonya
					Tarabunga}}}, \bibinfo {author} {\bibfnamefont {D.}~\bibnamefont {{Singh
					Bhakuni}}}, \bibinfo {author} {\bibfnamefont {M.}~\bibnamefont {{Dalmonte}}},
		\bibinfo {author} {\bibfnamefont {P.}~\bibnamefont {{Sierant}}},\ and\
		\bibinfo {author} {\bibfnamefont {X.}~\bibnamefont {{Turkeshi}}},\ }\bibfield
	{title} {\bibinfo {title} {{Universal Spreading of Nonstabilizerness and
				Quantum Transport}},\ }\href {https://doi.org/10.48550/arXiv.2506.12133}
	{\bibfield  {journal} {\bibinfo  {journal} {arXiv e-prints}\ ,\ \bibinfo
			{eid} {arXiv:2506.12133}} (\bibinfo {year} {2025})}\BibitemShut {NoStop}%
	\bibitem [{\citenamefont {{Martinez-Azcona}}\ \emph {et~al.}(2025)\citenamefont
		{{Martinez-Azcona}}, \citenamefont {{Sarkis}}, \citenamefont {{Tkatchenko}},\
		and\ \citenamefont {{Chenu}}}]{Chenu2025}%
	\BibitemOpen
	\bibfield  {author} {\bibinfo {author} {\bibfnamefont {P.}~\bibnamefont
			{{Martinez-Azcona}}}, \bibinfo {author} {\bibfnamefont {M.}~\bibnamefont
			{{Sarkis}}}, \bibinfo {author} {\bibfnamefont {A.}~\bibnamefont
			{{Tkatchenko}}},\ and\ \bibinfo {author} {\bibfnamefont {A.}~\bibnamefont
			{{Chenu}}},\ }\bibfield  {title} {\bibinfo {title} {{Magic Steady State
				Production: Non-Hermitian and Stochastic pathways}},\ }\href
	{https://doi.org/10.48550/arXiv.2507.08676} {\bibfield  {journal} {\bibinfo
			{journal} {arXiv e-prints}\ ,\ \bibinfo {eid} {arXiv:2507.08676}} (\bibinfo
		{year} {2025})}\BibitemShut {NoStop}%
	\bibitem [{\citenamefont {Cepollaro}\ \emph {et~al.}(2025)\citenamefont
		{Cepollaro}, \citenamefont {Akil}, \citenamefont {Cie{\'{s}}li{\'{n}}ski},
		\citenamefont {de~la Hamette},\ and\ \citenamefont {Brukner}}]{Brukner2025}%
	\BibitemOpen
	\bibfield  {author} {\bibinfo {author} {\bibfnamefont {C.}~\bibnamefont
			{Cepollaro}}, \bibinfo {author} {\bibfnamefont {A.}~\bibnamefont {Akil}},
		\bibinfo {author} {\bibfnamefont {P.}~\bibnamefont {Cie{\'{s}}li{\'{n}}ski}},
		\bibinfo {author} {\bibfnamefont {A.-C.}\ \bibnamefont {de~la Hamette}},\
		and\ \bibinfo {author} {\bibfnamefont {{\v{C}}.}~\bibnamefont {Brukner}},\
	}\bibfield  {title} {\bibinfo {title} {{Sum of Entanglement and Subsystem
				Coherence Is Invariant under Quantum Reference Frame Transformations}},\
	}\href {https://doi.org/10.1103/h6b3-y4vt} {\bibfield  {journal} {\bibinfo
			{journal} {Physical Review Letters}\ }\textbf {\bibinfo {volume} {135}},\
		\bibinfo {pages} {010201} (\bibinfo {year} {2025})}\BibitemShut {NoStop}%
\end{thebibliography}
\end{document}